\documentclass[preprint2]{aastex}

\newcommand{\GJ}{GJ$\,$876}
\newcommand{\au}{{\rm\,AU}}
\newcommand{\degperday}{^\circ {\rm\,day}^{-1}}
\newcommand{\yr}{{\rm\,yr}}
\newcommand{\s}{\scriptscriptstyle}

\begin{document}

\thispagestyle{empty}

\title{\vskip-10pt Dynamics and Origin of the 2:1 Orbital Resonances of\\
       the GJ$\,$876 Planets}
\author{Man Hoi Lee and S. J. Peale}
\affil{Department of Physics, University of California,
       Santa Barbara, CA 93106}

\begin{abstract}
The discovery by Marcy et al. (2001) of two planets in 2:1 orbital
resonance about the star \GJ\ has been supplemented by a dynamical fit to
the data by Laughlin \& Chambers (2001) which places the planets in coplanar
orbits deep in three resonances at the 2:1 mean-motion commensurability.
The selection of this almost singular state by the dynamical fit
means that the resonances are almost certainly real, and with the small
amplitudes of libration of the resonance variables, indefinitely
stable.
Several unusual properties of the 2:1 resonances are revealed by the \GJ\
system.
The libration of both lowest order mean-motion resonance variables and the
secular resonance variable, $\theta_1 = \lambda_1 - 2\lambda_2 + \varpi_1$,
$\theta_2 = \lambda_1 - 2\lambda_2 + \varpi_2$, and $\theta_3 = \varpi_1
- \varpi_2$, about $0^\circ$ (where $\lambda_{1,2}$ are the mean longitudes
of the inner and outer planet and $\varpi_{1,2}$ are the longitudes of
periapse) differs from the familiar geometry of the Io-Europa pair,
where $\theta_2$ and $\theta_3$ librate about $180^\circ$.
By considering the condition that ${\dot\varpi}_1 = {\dot\varpi}_2$ for
stable simultaneous librations of $\theta_1$ and $\theta_2$,
we show that the \GJ\ geometry results because of the large orbital
eccentricities $e_i$, whereas the very small eccentricities in the
Io-Europa system lead to the latter's geometry.
Surprisingly, the \GJ\ configuration, with $\theta_1$, $\theta_2$, and
$\theta_3$ all librating, remains stable for $e_1$ up to 0.86 and
for amplitude of libration of $\theta_1$ approaching $45^\circ$ with the
current eccentricities --- further supporting the indefinite stability of
the existing system.

Any process that drives originally widely separated orbits toward each
other could result in capture into the observed resonances at the 2:1
commensurability.
We find that forced inward migration of the outer planet of the \GJ\
system results in certain capture into the observed resonances if
initially $e_1\la 0.06$ and $e_2\la 0.03$ and the migration rate
$|\dot a_2/a_2| \la 3\times 10^{-2}(a_2/\au)^{-3/2}\yr^{-1}$.
Larger eccentricities lead to likely capture into higher order resonances
before the 2:1 commensurability is reached.
The planets are sufficiently massive to open gaps in the nebular disk
surrounding the young \GJ\ and to clear the disk material between them,
and the resulting planet-nebular interaction typically forces the outer
planet to migrate inward on the disk viscous time scale, whose inverse is
about three orders of magnitude less than the above upper bound on
$|{\dot a}_2/a_2|$ for certain capture.
If there is no eccentricity damping,
eccentricity growth is rapid with continued migration within the resonance,
with $e_i$ exceeding the observed values after a further reduction in
the semi-major axes $a_i$ of only 7\%.
With eccentricity damping ${\dot e}_i/e_i=-K|{\dot a}_i/a_i|$,
the eccentricities reach equilibrium values that remain constant for
arbitrarily long migration within the resonances.
The equilibrium eccentricities are close to the observed eccentricities for
$K\approx 100$ if there is migration and damping of the outer planet only,
but for $K\approx 10$ if there is also migration and damping of the inner
planet.
This result is independent of the magnitude or functional form of the
migration rate ${\dot a}_i$ as long as ${\dot e}_i/e_i =
-K |{\dot a}_i/a_i|$.
Although existing analytic estimates of the effects of planet-nebula
interaction are consistent with this form of eccentricity damping for certain
disk parameter values,
it is as yet unclear that such interaction can produce the large
value of $K$ required to obtain the observed eccentricities.
The alternative eccentricity damping by tidal dissipation within the star
or the planets is completely negligible, so the observed dynamical
properties of the \GJ\ system may require an unlikely fine tuning of the
time of resonance capture to be near the end of the nebula lifetime.
\end{abstract}

\section{INTRODUCTION}

Marcy et al. (2001) have discovered two planets about the nearby M dwarf
star \GJ.
A preliminary fit of the stellar radial velocity (RV) variations due to
two unperturbed Kepler orbits implies that the orbital periods of the two
planets are nearly in the ratio 2:1.
This resonance is an analog to the orbital resonances among the satellites
of Jupiter and Saturn (e.g., Peale 1999), but it is the first to be
discovered among extrasolar planetary systems.
Table~\ref{table1} shows the system parameters from the Marcy et al.
analysis based on data taken at both the Keck and Lick Observatories and
an adopted stellar mass of $0.32 M_\odot$.
The orbital periods are approximately 30 and 60 days.
The reduced chi-square statistic of $\chi_\nu^2=1.88$ indicates that the
two-Kepler system is an adequate fit to the data. However, the rather
large minimum masses of the planets of 0.56 and 1.89 $M_{\s J}$ (Jupiter
masses) mean that the two-Kepler fit may not be a good determination
of the system characteristics because the large mutual perturbations
will ensure that the orbits deviate from Kepler orbits. In fact,
substitution of the Marcy et al. parameters from Table~\ref{table1} (with
$\sin{i} = 1$ for both planets, where $i$ is the inclination of the orbital
plane to the plane of the sky) into a calculation of the perturbed
orbital motions beginning at the specified initial epoch leads
to large variations in the orbital elements that were assumed constant
in the fit.   On the other hand, projecting the orbital parameters in
Table~\ref{table1} to those at a different epoch within the time span of
the data
set leads to much less variations in the orbital elements and apparent
long term stability of the system (Marcy et al. 2001). For some
choices of epoch, the system is not in the 2:1 resonance, and it
eventually becomes unstable (Laughlin \& Chambers 2001). Clearly, a
dynamical fit of system characteristics to the RV observations is
necessary to constrain the parameters that define an RV curve for \GJ.

The publication of the data set by Marcy et al. (2001) allowed Laughlin
\& Chambers (2001) to perform such a fit. They assumed that the two
planets are on coplanar orbits, and used two methods to minimize
$\chi_\nu^2$ as a function of the initial system parameters.
Starting with the 2-Kepler fit, a Levenberg-Marquardt minimization scheme
driving an $N$-body integrator was used to find a {\it local} minimum in
$\chi_\nu^2$. A second method uses a genetic algorithm combined with
a simple model for the variations of the orbital elements due to the
2:1 orbital resonance to search for the {\it global} minimum in
$\chi_\nu^2$.  Both
methods converged to similar solutions. The dynamical fits are sensitive to
$\sin{i}$ since the determined masses (and perturbations) grow
as $i$ is decreased.  The best-fit solutions obtained using the
Levenberg-Marquardt method are included in Table~\ref{table1} for both the
Keck$+$Lick data and Keck data alone.
The Keck data is much more accurate than the Lick data and the solution for
this data alone has a minimum $\chi_\nu^2=1.59$ with an rms scatter of
$6.86\,{\rm m}\,{\rm s}^{-1}$ for $\sin{i} = 0.55$.
Although the Lick data is less accurate than the Keck data, the longer time
span of observation that includes the Lick data may give the better
solution.
The solution for the Keck$+$Lick data has a minimum $\chi_\nu^2=1.46$
with an rms scatter of $13.95\,{\rm m}\,{\rm s}^{-1}$ for $\sin{i} = 0.78$,
but the $\chi_\nu^2$ minimum is broad, with several nearby solutions giving
nearly as good a fit.
The conclusion of Laughlin \& Chambers is that the broad minimum in
$\chi_\nu^2$ allows probable values of $0.5<\sin{i}<0.8$.
(Rivera \& Lissauer 2001 examined mutually inclined orbits and got
similarly small $\chi_\nu^2$ values for several different configurations,
but they confirmed the Laughlin \& Chambers results for coplanar orbits.)

\begin{figure}[h]
\plotone{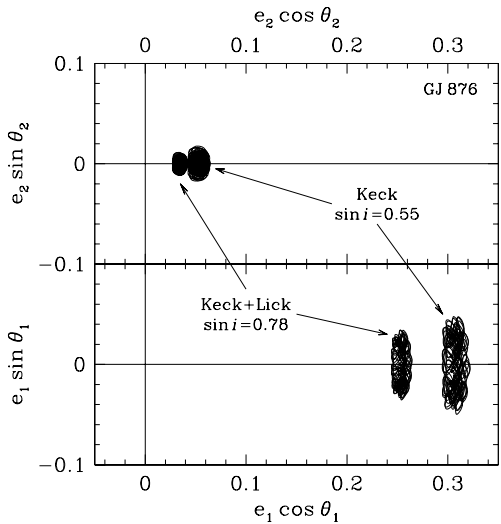}
\caption{
Small amplitude librations of the two 2:1 mean-motion resonance variables,
$\theta_1 = \lambda_1 - 2\lambda_2 + \varpi_1$ and $\theta_2 = \lambda_1 -
2\lambda_2 + \varpi_2$, about $0^\circ$ for the \GJ\ planets.
Trajectories for 3100 days (the average periapse precession period for
both planets) are shown in plots of $e_j \sin\theta_j$ versus
$e_j \cos\theta_j$ for the Laughlin \& Chambers (2001) best-fit solutions
to the Keck data alone and the combined Keck and Lick data.
\label{figure1}}
\end{figure}

Figure~\ref{figure1} shows the trajectories of the motions for 3100 days
(the average periapse precession period for both planets) for the two
Laughlin-Chambers best-fit solutions in plots of $e_j \sin\theta_j$ versus
$e_j \cos\theta_j$, where $e_j$ is the eccentricity of the $j$th planet
(with $j=1$ and $2$ for the inner and outer planets, respectively),
\begin{eqnarray}
\begin{array}{rl}
\theta_1 &= \lambda_1 - 2 \lambda_2 + \varpi_1 ,\\
\theta_2 &= \lambda_1 - 2 \lambda_2 + \varpi_2 ,
\end{array}
\end{eqnarray}
are the two lowest order, eccentricity-type mean-motion resonance variables
at the 2:1 commensurability,
and $\lambda_j$ and $\varpi_j$ are the mean longitude and longitude
of periapse of the $j$th planet.
The two Laughlin-Chambers solutions place the system deep in both
mean-motion resonances, with $\theta_1$ and $\theta_2$ librating about
$0^\circ$ with remarkably small amplitudes.
The simultaneous librations of $\theta_1$ and $\theta_2$ about $0^\circ$
mean that the secular resonance variable
\begin{equation}
\theta_3 = \varpi_1 - \varpi_2 = \theta_1 - \theta_2
\end{equation}
also librates about $0^\circ$.
Although the parameters will be more tightly defined
as more data is acquired, the 
almost singular nature of the low amplitude librations indicates that
the real system is most likely indeed locked in multi-resonance
librations at the 2:1 mean-motion commensurability.  This
means that the line of apsides of the two orbits are nearly
aligned with conjunctions of the two planets always occurring very
close to the periapse longitudes.  The resonance configuration ensures
that the planets can never make close approaches in spite of their
large masses, and barring external perturbations or an unreasonably
high dissipation of tidal energy in the planets, the system should be 
stable for the lifetime of the star.  In fact, we find that
the librations are stable even for amplitudes of the inner planet resonance
variable $\theta_1$ approaching $45^\circ$ if we induce larger amplitudes of
libration of the resonance variables by changing the initial value of the
mean anomaly of the inner planet.  The existence of the mean-motion
resonances means that the assumption that the orbits are
nearly coplanar should be correct, and such coplanarity is consistent
with the accretion of planetary bodies in a nebular disk surrounding
the forming star.  We shall assume that the orbits are coplanar
hereinafter. 
Furthermore, there is not much to distinguish the two Laughlin-Chambers
solutions in Figure~\ref{figure1}, and we shall assume the parameters
appropriate to the solution based on both the Keck and Lick data.

\begin{figure}[t]
\plotone{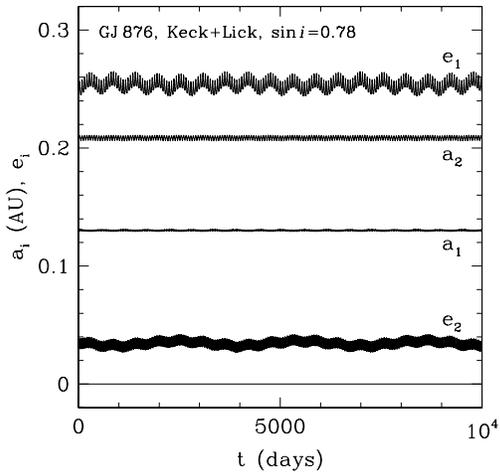}
\caption{
Variations in the semi-major axes, $a_1$ and $a_2$, and eccentricities,
$e_1$ and $e_2$, of the \GJ\ planets for the Laughlin-Chambers
Keck$+$Lick solution.
The small amplitude librations of the resonance variables ensure that the
semi-major axes and eccentricities have little variation.
\label{figure2}}
\end{figure}

Figure~\ref{figure2} shows the variations in the semi-major axes and
eccentricities of both planets for $10^4$ days for the Laughlin-Chambers
Keck$+$Lick solution.
The small amplitude librations of the resonance variables about $0^\circ$
ensure that both the eccentricities and semi-major axes have little
variation in spite of the large mutual perturbations.
This is why the 2-Kepler fit of Marcy et al. (2001) could produce a fit
that was not too bad.

The fact that both 2:1 mean-motion resonance variables librate about
$0^\circ$ in Figure~\ref{figure1} contrasts with the geometry of
the Io-Europa 2:1 orbital resonances at Jupiter, where the resonance
variable involving Io's longitude of periapse, $\theta_1$, librates about
$0^\circ$ but the resonance variable involving Europa's longitude of
periapse, $\theta_2$, librates about $180^\circ$ (e.g., Peale 1999).
This means that in the Io-Europa case the lines of 
apsides are anti-aligned, with the periapses $180^\circ$ apart (i.e.,
$\theta_3$ librates about $180^\circ$).
Conjunctions occur when Io is near periapse and Europa is near
apoapse.
In \S 2 we explain why the \GJ\ and Io-Europa systems have different
resonance configurations by considering a condition for stable
simultaneous librations of the two mean-motion resonance variables.
In particular, the longitudes of periapse should precess in the same
direction at the same average rate, so that the relative alignment of the
lines of apsides is maintained.
In the Io-Europa case, the orbital eccentricities are sufficiently small
that the precession rate $d\varpi_i/dt$ is dominated by a single
term lowest order in eccentricities and containing $\cos\theta_i$.
The overall sign of the coefficient for this term is $<0$ for $i = 1$
(inner satellite) but $>0$ for $i = 2$ (outer satellite), and
retrograde precessions of both satellites require $\theta_1$ to
librate about $0^\circ$ and $\theta_2$ to librate about $180^\circ$.
In the case of \GJ, the eccentricities are sufficiently large that
there are large contributions to the precession rates from higher order
terms whose cosine arguments are linear combinations of the resonance
variables.
Summing over all contributing terms, coincident retrograde precessions of
the \GJ\ planets require both $\theta_1$ and $\theta_2$ to librate about
$0^\circ$.

If the orbits of the two planets about \GJ\ were originally much further
apart, with the ratio of their mean motions considerably greater
than 2:1, any process that drives the orbits toward each other could lead
to capture into the 2:1 orbital resonances that we observe.
In \S 3 we describe a particular migration process due to the gravitational
interaction between the planets and the nebular disk surrounding the young
\GJ.
We note the simulations by Bryden et al. (2000) and Kley (2000), which
show that two planets that are massive enough to open gaps individually in
the gas disk can rather quickly clear out the disk material between them,
if they are not separated too far.
Disk material outside the outer planet exerts torques on the planet that
are not opposed by disk material on the inside, and the outer planet
migrates toward the star on the disk viscous time scale.
Any disk material left on the inside of the inner planet exerts torques on
the inner planet that push it away from the star.
Thus the condition of approaching orbits necessary to form the resonances
is established.
We discuss the effect of planet-nebula interaction on orbital
eccentricities and note that the \GJ\ system is in the interesting regime
where it is uncertain whether eccentricity damping or growth is expected
because of the rather large planet-star mass ratio of the outer planet.
(Although the mass of the outer planet, $M_2$, for the Laughlin-Chambers
Keck$+$Lick solution is only $2.40 M_{\s J}$, $M_2/M_0 = 7.17 \times
10^{-3}$ because the stellar mass $M_0 = 0.32 M_\odot$.)

In \S 4 we present the results of a series of numerical orbit integrations
in which the orbits of the \GJ\ planets, initially far from the 2:1
mean-motion commensurability, are forced to approach each other.
(The numerical methods are described in the Appendix.)
We show that capture into the two 2:1 mean-motion resonances and the
secular resonance is certain if the orbital eccentricities start
reasonably small and the rate of migration is not too fast.
If there is no eccentricity damping, the eccentricities of both planets
increase rapidly after resonance capture and exceed the observed values
after a very short migration of the resonantly locked planets.
Eccentricity damping of the form ${\dot e}_i/e_i \propto {\dot a}_i/a_i$,
where a dot over a symbol denotes $d/dt$ and ${\dot a}_i$ is the forced
migration rate, leads to a termination in the eccentricity growth, and the
eccentricities reach equilibrium values that remain constant for
arbitrarily long migration in the resonances.
We find that significant eccentricity damping with $|{\dot e}_i/e_i| \gg
|{\dot a}_i/a_i|$ is required to produce the observed eccentricities of
the \GJ\ system.
In \S 5.1 we show that alternative damping of the eccentricities by tidal
dissipation within the star or planets is insignificant, and in \S 5.2 we
discuss other related studies.
Our conclusions are summarized in \S 6.

\section{COMPARISON WITH THE\\ IO-EUROPA SYSTEM}

As shown in Figure~\ref{figure1}, the \GJ\ system has both lowest order,
eccentricity-type mean-motion resonance variables at the 2:1
commensurability, $\theta_1 = \lambda_1 - 2 \lambda_2 + \varpi_1$ and
$\theta_2 = \lambda_1 - 2 \lambda_2 + \varpi_2$, and hence the secular
resonance variable, $\theta_3 = \varpi_1 - \varpi_2$, librating about
$0^\circ$.
This resonance configuration is different from that of Io and Europa,
where $\theta_2$ and $\theta_3$ librate about $180^\circ$.
It should be noted that the differences are not due to the additional
resonances involving Ganymede in the Io-Europa case.
In the scenario where the resonances among the inner three Galilean
satellites of Jupiter are assembled by differential tidal expansion of the
orbits (Yoder 1979; Yoder \& Peale 1981), Io is driven out most rapidly and
the resonance variables involving Io and Europa only are captured into
libration first.
These resonance variables have the same centers of libration before and
after the 2:1 commensurability between Europa and Ganymede is encountered.
The differences in the resonance configurations of \GJ\ and Io-Europa
are instead due to the magnitudes of the eccentricities involved and can be
understood from a condition for stable simultaneous librations of the
two mean-motion resonance variables.

This condition for stable simultaneous librations of $\theta_1$
and $\theta_2$ is that
the longitudes of periapse, $\varpi_1$ and $\varpi_2$, on average
precess at the same rate.
For coplanar orbits, the equations for the variation of $\varpi_i$, $a_i$,
and $e_i$ in Jacobi coordinates are (e.g., Peale 1986)
\begin{eqnarray}
{d\varpi_i \over dt} &=& -{\sqrt{1-e_i^2} \over M'_ie_i\sqrt{\mu_i a_i}}\,
                         {\partial H \over \partial e_i},
\label{precess} \\ & & \nonumber \\
{da_i \over dt}      &=& -{2 \over M'_i}\sqrt{a_i \over \mu_i}\,
                         {\partial H \over \partial \lambda_i},
\label{dadt} \\ & & \nonumber \\
{de_i \over dt}      &=& {\sqrt{1-e_i^2} \over M'_ie_i\sqrt{\mu_i a_i}}\,
                         {\partial H \over \partial \varpi_i}
\nonumber \\
                     &&    -{(1-e_i^2) - \sqrt{1-e_i^2}
                               \over M'_ie_i\sqrt{\mu_i a_i}}\,
                         {\partial H \over \partial \lambda_i},
\label{dedt}
\end{eqnarray}
where $M'_i = M_i \sum_{k=0}^{i-1} M_k/\sum_{k=0}^i M_k$,
$\mu_i = G M_0 M_i/M'_i$, $M_0$ is the mass of the star (or Jupiter),
$M_i$ is the mass of the $i$th planet (or satellite), the Hamiltonian
\begin{equation}
H = -\sum_{i=1}^2 {GM_0 M_i \over 2a_i} + \Phi,
\end{equation}
and the disturbing potential
\begin{equation}
\Phi = -{G M_1 M_2 \over r_{12}}-G M_0 M_2
       \left({1 \over r_{02}}-{1 \over r_2}\right) .
\end{equation}
If we neglect terms of order $(M_1/M_0)^2$ and higher and assume that
$a_1 < a_2$, $\Phi$ can be expanded to the form
\begin{eqnarray}
\Phi &=& -{G M_1 M_2 \over a_2} \sum C^j_{k \ell m n}(\beta) \nonumber \\
     & & \qquad \times e_1^{|k| + 2m} e_2^{|\ell|+2n} \cos{\phi_{j k \ell}},
\label{phiexp}
\end{eqnarray}
where
\begin{equation}
\phi_{j k \ell} =
(j - k) \lambda_1 - (j - \ell) \lambda_2 + k \varpi_1 - \ell \varpi_2 ,
\end{equation}
and the summation is over the range $0 \le j \le \infty$,
$-\infty \le k,\ell \le \infty$, and $0 \le m,n \le \infty$.
The coefficient $C^j_{k \ell m n}(\beta)$, where $\beta = a_1/a_2$, can
be written in terms of the Laplace coefficient $b_{1/2}^j(\beta)$ and its
derivatives (Brouwer \& Clemence 1961).
For a given cosine argument $\phi_{j k \ell}$, the term lowest order in
eccentricities is of order $e_1^{|k|} e_2^{|\ell|}$.

Within the 2:1 mean-motion resonances and the secular resonance,
the perturbations are dominated by the terms whose arguments are nearly
fixed because of the resonances, i.e., those terms involving the
resonance variables $\theta_1$, $\theta_2$, $\theta_3$,
and their linear combinations.
Since the terms in equation (\ref{dadt}) for $da_i/dt$ and equation
(\ref{dedt}) for $de_i/dt$ are proportional to
$\partial H/\partial \lambda_i$ and $\partial H/\partial \varpi_i$ and
$\lambda_i$ and $\varpi_i$ appear in the cosine arguments only,
the cosines are changed into sines and
there is no secular change in $a_i$ and the forced $e_i$ if the resonance
variables $\theta_1$, $\theta_2$, and $\theta_3$ librate about either
$0^\circ$ or $180^\circ$.
(Note, however, that if the eccentricities are not small and $da_i/dt$ and
$de_i/dt$ are not dominated by a single term, there is also the possibility
that the sums of all contributing terms are zero for resonance variables
having values other than $0^\circ$ and $180^\circ$.)

We have calculated the contributions to the precession rate
$d\varpi_i/dt$ from secular terms in equation (\ref{phiexp}) with argument
of the form $k \theta_3$ (including the $k = 0$ non-resonant secular term),
up to fourth order in eccentricities, and from mean-motion resonance terms
with argument of the
form $k \theta_1 - \ell \theta_2$ ($k \ne \ell$), up to third order
in eccentricities, and they are listed in Table~\ref{table2}.
We set the resonance variables to their average values, i.e., the
libration center values.
In addition, we ignore the small deviation of $\beta$ from $2^{-2/3}$
(since the commensurability is not exact) and evaluate
$C^j_{k \ell m n}(\beta)$ at $\beta = 2^{-2/3}$.
For Io and Europa, we adopt $e_1 = 0.0026$ and $e_2 = 0.0013$, which are
the equilibrium eccentricities before the resonances with Ganymede are
encountered in the tidal scenario (Yoder \& Peale 1981).
The numerical values in Table~\ref{table2} are for the current semi-major
axis of Io, but they can be scaled to other values of the semi-major
axis since the precession rates are proportional to $a_1^{-3/2}$.
As we can see in Table~\ref{table2},
because the orbital eccentricities of Io and Europa are small,
the precession rate $d\varpi_i/dt$ is dominated by a single term lowest
order in the eccentricities.
This is the term with argument $\theta_i$ and proportional to $e_i$ in
the disturbing potential (eq.~[\ref{phiexp}]), and whose contribution to
$d\varpi_i/dt$ is proportional to $1/e_i$ (eq.~[\ref{precess}]).
Since the coefficients of the $e_1 \cos\theta_1$ and $e_2 \cos\theta_2$
terms are $C^2_{1000}=-1.19$ and $C^1_{0-100}=+0.43$, respectively,
retrograde precessions of both Io and Europa require $\theta_1 = 0^\circ$
and $\theta_2 = 180^\circ$.
Furthermore, the requirement that the regression rates are identical implies 
a simple relationship between the eccentricities:
\begin{equation}
\beta n_1\frac{M_2}{M_0}\frac{C_{1000}^2}{e_1} + \dot\varpi_{{\rm sec},1} =
- n_2\frac{M_1}{M_0}\frac{C_{0-100}^1}{e_2} +\dot\varpi_{{\rm sec},2},
\label{eq:e1e2}
\end{equation}
where $n_i$ is the mean motion and the secular motion
$\dot\varpi_{{\rm sec},i}$ includes contributions from secular terms in the
disturbing potential and the additional secular motion induced by the
oblateness of Jupiter.
As we can see in Table~\ref{table2}, the contributions from the secular
terms in the disturbing potential and even the secular motion induced by
the oblateness of Jupiter (which is $+0.12\degperday$ for Io's orbit)
are small compared with the $\sim -1.5\degperday$ precession induced
by the first order mean-motion resonance terms with arguments $\theta_1$ and
$\theta_2$.
Thus $e_1/e_2\approx -\beta^{-1/2}(C_{1000}^2/C_{0-100}^1)
(M_2/M_1)=1.9$, where the last equality is for the masses of Io and Europa.

For \GJ, we know from numerical orbit integration of the
Laughlin-Chambers Keck$+$Lick solution that on average
$d\varpi_1/dt = d\varpi_2/dt = -0.116\degperday$.
The numerical values in Table~\ref{table2} are obtained using $e_1 = 0.255$,
$e_2 = 0.035$, and $a_1 = 0.130\au$, which are the average values
for the Laughlin-Chambers Keck$+$Lick solution (see Fig.~\ref{figure2}).
As we would expect from the analysis above for the Io-Europa case,
with $\theta_1 = \theta_2 = 0^\circ$, the contributions to $d\varpi_1/dt$
and $d\varpi_2/dt$ from the $e_1 \cos\theta_1$ and $e_2 \cos\theta_2$ terms,
respectively, have opposite signs.
However, because the eccentricities are large, these terms no longer
dominate the precession rates.
In particular, the largest contribution to $d\varpi_2/dt$ comes from terms
second order in eccentricities --- the largest of which being the
$e_1 e_2 \cos(\theta_1 + \theta_2)$ term, which is of order $e_1$($=0.255$)
lower than the $e_2 \cos\theta_2$ term, but whose coefficient is large and
negative ($C^3_{1-100} = -4.97$).
We can also see from Table~\ref{table2} that both $d\varpi_1/dt$ and
$d\varpi_2/dt$ converge only slowly with the order of the $k\theta_1 -
\ell\theta_2$ terms.
With the inclusion of terms up to $e^3$, the precession rates have the
correct, negative signs when $\theta_1 = \theta_2 = \theta_3 = 0^\circ$,
but the magnitude of the precession rate for the inner (outer) planet
is significantly larger (smaller) than that for both planets from numerical
orbit integration.
Based on the magnitudes and the alternating signs of the contributions from
the lower order terms, the expected contributions from the $e^4$ (and
higher order) terms are consistent with their bringing the results into
agreement with the actual precession rates.
There is no simple analytic expression relating $e_1$ to $e_2$ when the
eccentricities are large, but as in the small eccentricity limit, the
eccentricities are related by the requirement that the periapses precess
at the same rate.

Although the analysis in this section is able to explain why the \GJ\ and
Io-Europa systems have different resonance configurations, the slow
convergence of the series for moderate to large eccentricities means that
it has limited usefulness if one is interested in the more general question
of what stable configurations are possible for different periapse
precession rates and masses.
A practical way to investigate this latter question is through numerical
migration calculations like those in \S 4.1, which drive a system
through a sequence of configurations with different precession rates.
As we shall see, for systems with masses like those in \GJ, there are
stable configurations with $\theta_1$, $\theta_2$, and $\theta_3$
librating about $0^\circ$ for $0.15 \la e_1 \la 0.86$
(see Fig.~\ref{figure3}).
We have also found that configurations with resonance variables
librating about angles other than $0^\circ$ and $180^\circ$ are possible
when the masses are different from those in \GJ;
these configurations will be discussed in a subsequent paper.

\section{MIGRATION SCENARIO FOR ORIGIN OF RESONANCES}

We now turn to the question of the origin of the two 2:1 mean-motion
resonances and the secular resonance in the \GJ\ system.
A scenario for the origin of this resonance configuration is that it was
assembled by differential migration of orbits that were initially much
further apart.
We shall see in \S 4 that libration of all three resonance variables is
easily established by any process that drives the orbits toward each other.
Although the results presented in \S 4 are quite general, they will
be interpreted in the context of a particular differential migration
process, namely via the gravitational interaction between the planets and
the nebula from which they form.

It is generally accepted that planets form in a disk of gas and dust that
surrounds a young star.
Since the two planets around \GJ\ are of order Jupiter mass and are
presumably gas giants, they must have formed within the lifetime of the
gas disk.
If a planet is sufficiently massive, the torques exerted by the planet on
the gas disk can open an annular gap in the disk about the planet's orbit.
The conditions for gap formation are that the Roche radius of the planet,
$r_R = (M/3 M_0)^{1/3} a$, exceeds the scale height $H$ of
the disk, or equivalently,
\begin{equation}
M/M_0 \ga 3 (H/a)^3 = 3.75 \times 10^{-4} \left(H/a \over 0.05\right)^3,
\end{equation}
and that the viscous condition
\begin{eqnarray}
M/M_0 \!\!&\ga&\!\! 40 \nu/(\Omega a^2)=40 \alpha (H/a)^2 \nonumber\\
      \!\!&=&\!\!   4 \times 10^{-4} \left(\alpha \over 4\times 10^{-3}\right)
                    \left(H/a \over 0.05\right)^2 \nonumber\\
& &
\end{eqnarray}
is satisfied (e.g., Lin \& Papaloizou 1993).
In the above equations, $M$ and $M_0$ are the masses of the planet and the
star, $a$ is the semi-major axis of the planet's orbit, $\Omega$ is the angular
Kepler speed at $a$, and the kinematic viscosity $\nu$ is expressed using
the Shakura-Sunyaev $\alpha$ prescription: $\nu=\alpha H^2 \Omega$.
Since the planet-star mass ratios of the planets around \GJ\ are
$M_1/M_0 = 2.28 \times 10^{-3}$ and $M_2/M_0 = 7.17 \times
10^{-3}$, these planets are expected to open gaps individually during their
growth to their final masses if their orbits are sufficiently far apart,
unless $H/a \ga 0.09$ or $\alpha^{1/2} (H/a) \ga 7.5 \times 10^{-3}$.

The numerical values $\alpha = 4 \times 10^{-3}$ and $H/a = 0.05$ are
typical of models of protoplanetary disks (e.g., Bryden et al. 2000;
Kley 2000; Papaloizou et al. 2001).
Hartmann et al. (1998) have also inferred from observed properties of
T Tauri disks that $\alpha \sim 10^{-2}$.
The most promising source of an effective viscosity in accretion disks is
magnetohydrodynamic (MHD) turbulence initiated and sustained by the
magnetorotational instability, which is capable of producing an effective
$\alpha$ as large as $0.1$ (e.g., Stone et al. 2000).
However, protoplanetary disks are probably too weakly ionized for MHD
turbulence to develop fully, except at small radii ($\la 0.1\au$) and
possibly in a layer near the surface of the disk at larger radii
(Gammie 1996).
Another possible source of effective viscosity is damping of density waves
excited by many small (of order Earth mass) planets, which is capable of
producing an effective $\alpha \la 10^{-3}$ (Goodman \& Rafikov 2001).

Bryden et al. (2000) and Kley (2000) have performed hydrodynamic
simulations of a system consisting of a central star, a gas disk and two
planets.
The planets are massive enough ($M/M_0 \approx 10^{-3}$) to open gaps
individually, and the ratio of their initial semi-major axes is $a_1/a_2
\approx 1/2$.
They found that the planets clear out nearly all of the nebular material
between them in a few hundred orbital periods.
Then the torques exerted by the nebular material outside the outer planet
drives that planet toward the star, whereas any nebular material left on
the inside of the inner planet drives that planet away from the star.
The depletion of the inner disk means that the inner planet may not
move out very far, but the net effect is always to drive the orbits of two
massive planets toward each other.
The time scale on which the planets migrate is the disk viscous time scale,
whose inverse is (Ward 1997)
\begin{eqnarray}
\left|{\dot a} \over a\right|
\!\!&\approx&\!\! {3 \nu \over 2 a^2}
= {3 \over 2} \alpha (H/a)^2 \Omega \nonumber \\
\!\!&=&\!\! 5.3 \times 10^{-5} \left(\alpha \over 4\times 10^{-3}\right)
            \left(H/a \over 0.05\right)^2 \nonumber \\
\!\!&&\!\!  \times\left(M_0 \over 0.32 M_\odot\right)^{1/2}
            \left(a \over {\rm AU}\right)^{-3/2} \yr^{-1}, \nonumber \\
&& \label{vistime}
\end{eqnarray}
where a dot over a symbol denotes $d/dt$.

In addition to its effect on the semi-major axis, planet-nebula interaction
can also affect the orbital eccentricity of a planet (e.g., Goldreich \&
Tremaine 1980; Artymowicz 1992, 1993; Papaloizou et al. 2001).
A planet interacts with a disk in the vicinity of Lindblad and corotation
resonances.
The leading contribution to ${\dot a}$ is due to Lindblad resonances
with the $l = m$ Fourier components of the planet's perturbation potential.
(The indices $l$ and $m$ for the Fourier series in time and azimuthal
angle in this context should not be confused with $\ell$ and $m$ in
eq.~[\ref{phiexp}].)
For a planet orbiting in a disk gap, the so-called co-orbital
Lindblad and corotation resonances are not important, and the leading
contributions to ${\dot e}$ are damping due to $l = m \pm 1$ corotation
resonances and excitation due to $l = m - 1$ outer and $l = m + 1$ inner
Lindblad resonances.
The net effect of the corotation resonance damping and the Lindblad
resonance excitation depends on the distribution of the nebular material,
which is itself determined by the interaction with the planet.
If the gap is not too wide and many resonances of both types are present,
previous calculations indicate that there is net eccentricity damping.
For example, if we consider an outer disk of constant surface mass density
$\Sigma$ that extends radially from $a + \Delta$ to $\infty$, with
$\Delta \ll a$, then integration of equations (30) and (31) of Goldreich
\& Tremaine (1980) yields
\begin{eqnarray}
{{\dot e} \over e} &=& -0.116 \left(M \over M_0\right)
\left(\Sigma a^2 \over M_0\right) \left(a \over \Delta\right)^4 \Omega ,
\label{GTdedt} \\
& & \nonumber \\
{{\dot a} \over a} &=& -1.67 \left(M \over M_0\right)
\left(\Sigma a^2 \over M_0\right) \left(a \over \Delta\right)^3 \Omega .
\label{GTdadt}
\end{eqnarray}
(Note that the above equations are identical to eqs. [109] and [110] of
Goldreich \& Tremaine where they set $\Delta = 2 a/3 m_{\rm max}$.)

In the case of \GJ, since the planets clear out nearly all of the
nebular material between them and the inner disk is likely to be depleted,
the dominant resonant interactions are expected to be those between the
outer disk and the outer planet.
However, the estimate (\ref{GTdedt}) for $\dot e$ is not adequate because,
in addition to the fact that the mass distribution near the inner edge of
the outer disk is not modeled properly,
the outer planet is sufficiently massive that only a few low-$m$
resonances are likely to be present in the disk.
In fact, if the outer planet is able to open a gap out to the 2:1
commensurability, there would be no corotation resonances and only one
Lindblad resonance with $l = m-1 = 1$ at the 3:1 commensurability, and
there would be eccentricity growth instead of damping (Artymowicz 1992).
(In addition, the disk can become eccentric and the growth of the planet's
orbital eccentricity can be enhanced by the interaction with the eccentric
disk, if the planetary mass is comparable to a characteristic mass of the
disk; Papaloizou et al. 2001.)
From a comparison of the resonant and viscous torques,
an approximate condition for opening a gap out to the 2:1 commensurability
is (Artymowicz 1992; Lin \& Papaloizou 1993)
\begin{eqnarray}
M/M_0 \!\!&\ga&\!\! 2.8 \alpha^{1/2} (H/a) \nonumber \\
\!\!&=&\!\! 8.9 \times 10^{-3} \left(\alpha \over 4\times 10^{-3}\right)^{1/2}
            \left(H/a \over 0.05\right) . \nonumber \\
&& \label{2to1cond}
\end{eqnarray}
Since the planet-star mass ratio of the outer planet around \GJ\ is
$M_2/M_0 = 7.17 \times 10^{-3}$ and close to the critical value in
equation (\ref{2to1cond}), without knowing the exact values of the disk
parameters, it is unclear whether one should expect eccentricity damping or
growth.
However, as we shall see in \S 4, significant eccentricity
{\it damping} with $|{\dot e}/e| \gg |{\dot a}/a|$ is required to produce
the observed eccentricities of the \GJ\ system, unless the migration
after resonance capture is severely limited.
Therefore, at least the condition (\ref{2to1cond}) must not be satisfied,
implying that
\begin{eqnarray}
\alpha^{1/2} (H/a) \!\!&\ga&\!\! 0.36 M_2/M_0 \nonumber \\
\!\!&=&\!\! 2.6 \times 10^{-3} \left( M_2/M_0 \over 7.17 \times 10^{-3}\right)
\nonumber \\
&& \label{2to1condII}
\end{eqnarray}
for the outer disk of the young \GJ\ system.
A more detailed analysis using hydrodynamic simulations is necessary to
determine whether sufficient eccentricity damping can be produced by such
an outer disk.

\section{NUMERICAL RESULTS FOR MIGRATION SCENARIO}

In this section we present the results of a series of numerical orbit
integrations designed to determine the conditions under which the dynamical
properties of the current \GJ\ system could be produced by any process
(such as the planet-nebular interaction discussed in \S 3) that drives
the orbits of the two planets toward each other.
We consider a system consisting of a central star and two planets, where
the stellar and planetary masses are those for the Laughlin-Chambers
Keck$+$Lick solution (Table~\ref{table1}).
Unless stated otherwise, the planets are initially on coplanar, circular
orbits, with the mean longitudes differing by $180^\circ$ and the ratio of
the semi-major axes $a_1/a_2 = 1/2$, far from the 2:1 mean-motion
commensurability.

In addition to the mutual gravitational interactions of the star and the
planets, we force the osculating semi-major axis of planet $i$ to migrate
at a rate ${\dot a}_i$.
In most cases, we assume that only the outer planet is forced to migrate
inward, with a migration rate of the form ${\dot a}_2/a_2 =$ constant.
The effects of adopting a more general form of the migration rate
(e.g., ${\dot a}_2/a_2$ being a function of $a_2$) or forcing the inner
planet to migrate outward are discussed in \S 4.2.
For the calculations in \S 4.2, we also damp the osculating eccentricity at
a rate ${\dot e}_i$.
We adopt a damping rate of the form ${\dot e}_i/e_i = -K |{\dot a}_i/a_i|$,
where $K$ is a positive constant.
As we shall see, this form of eccentricity damping has the convenient
property that the eccentricities reach equilibrium values after capture
into the 2:1 resonances.
We note that the relation ${\dot e}/e \propto {\dot a}/a$ is valid for the
simple estimates (\ref{GTdedt}) and (\ref{GTdadt}) for the interaction
of a massive planet with an outer disk if $\Delta/a \approx$ constant
and that the same relation between the damping and migration rates also
holds for a planet that is too small to open a gap in a gas disk with
constant $H/a$ and undergoes the so-called type I migration
(Artymowicz 1993; Ward 1997).

The numerical orbit integrations were performed using the symplectic
integrator SyMBA (Duncan, Levison, \& Lee 1998) modified to include the
orbital migration and eccentricity damping terms.
SyMBA is based on a variant of the Wisdom-Holman (1991) method
and employs a multiple time step technique to handle close encounters.
(The latter feature is not essential for the integrations presented here.)
We confirmed the results by repeating some of the integrations using a
Bulirsch-Stoer integrator that has also been modified to include migration
and damping.
We also checked that both modified integrators give the correct exponential
decays in $a$ and $e$ when they are used to integrate the orbit of a single
planet with ${\dot a}/a =$ constant and ${\dot e}/e =$ constant.
We describe how the integrators were modified in the Appendix.

\subsection{Migration without Eccentricity Damping}

We consider first inward migration of the outer planet without eccentricity
damping.
In Figure~\ref{figure3} we show the results of a calculation with $a_1 =
0.5\au$ and $a_2 = 1.0\au$ initially and ${\dot a}_2/a_2 = -5 \times 10^{-5}
\yr^{-1}$.
The migration rate is consistent with that due to planet-nebular interaction
at $\sim 1\au$ (eq.~[\ref{vistime}]).
Figure~\ref{figure3}{\it a} shows the time evolution of the semi-major axes
and eccentricities, and Figure~\ref{figure3}{\it b} shows the evolution of
the two 2:1 mean-motion resonance variables, $\theta_1$ and $\theta_2$, in
plots of $e_i \sin\theta_i$ versus $e_i \cos\theta_i$.
Initially, only the outer planet migrates inward at the prescribed rate.
When the 2:1 mean-motion commensurability is encountered, both $\theta_1$
and $\theta_2$ (and hence the secular resonance variable $\theta_3$) are
captured into libration about $0^\circ$.
We do not see libration of $\theta_2$ and $\theta_3$ about $180^\circ$,
which is expected for small eccentricities (see \S 2), because the
planetary masses are so large that fairly large eccentricities (with $e_1$
reaching about $0.15$) are generated before the 2:1 commensurability is
encountered.
Because of the forced migration, the centers of libration are actually
slightly offset from $0^\circ$.
The resulting resonant interaction slows down the migration of the outer
planet and forces the inner planet to migrate inward while keeping the
ratio of the semi-major axes nearly constant ($a_1/a_2 \approx 2^{-2/3}$);
it also causes the eccentricities to increase rapidly.
The centers of libration remain near $0^\circ$ and the amplitudes of
libration remain small as the eccentricities increase.

\begin{figure}[t]
\plotone{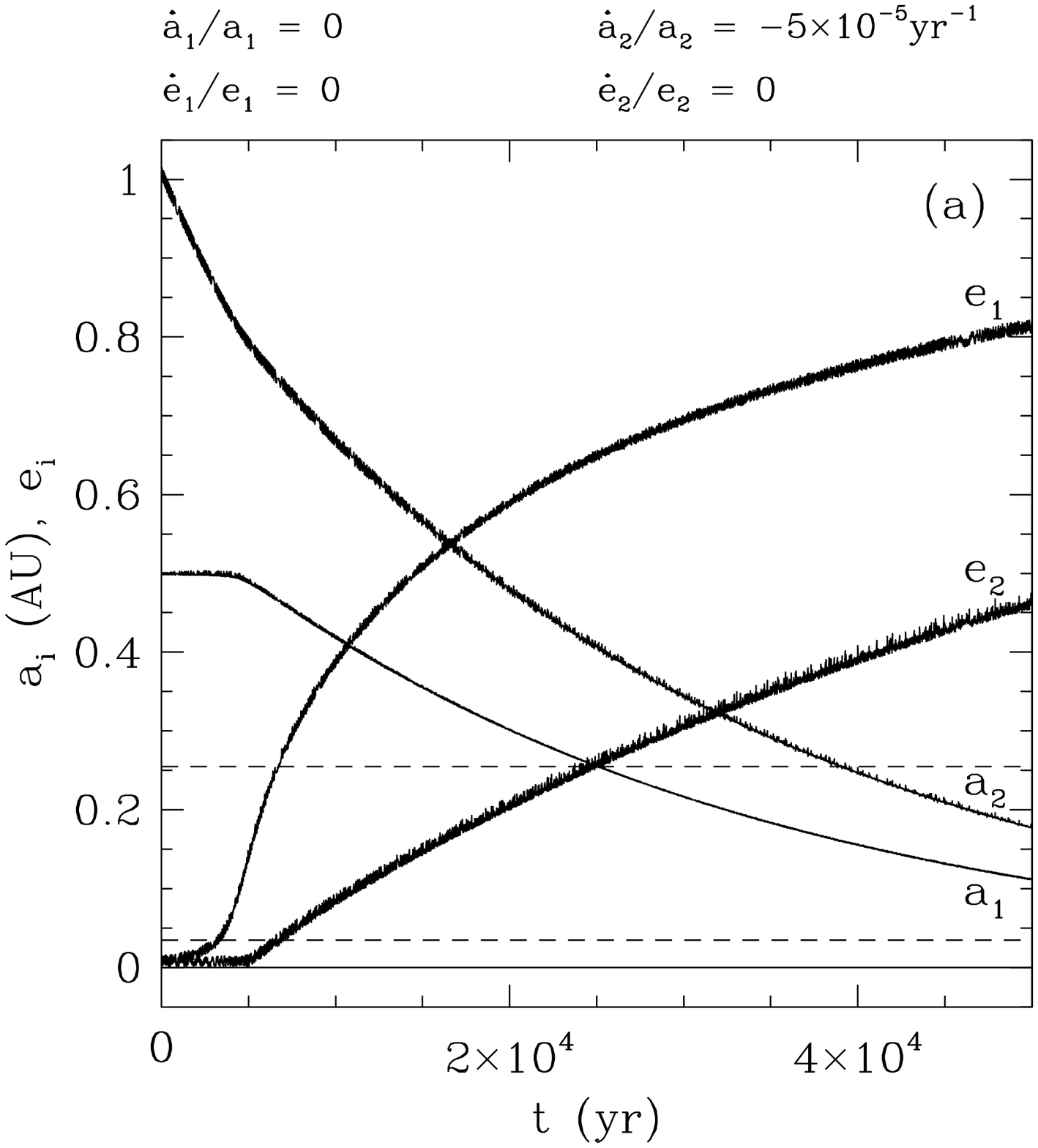}
\end{figure}
\begin{figure}[h]
\plotone{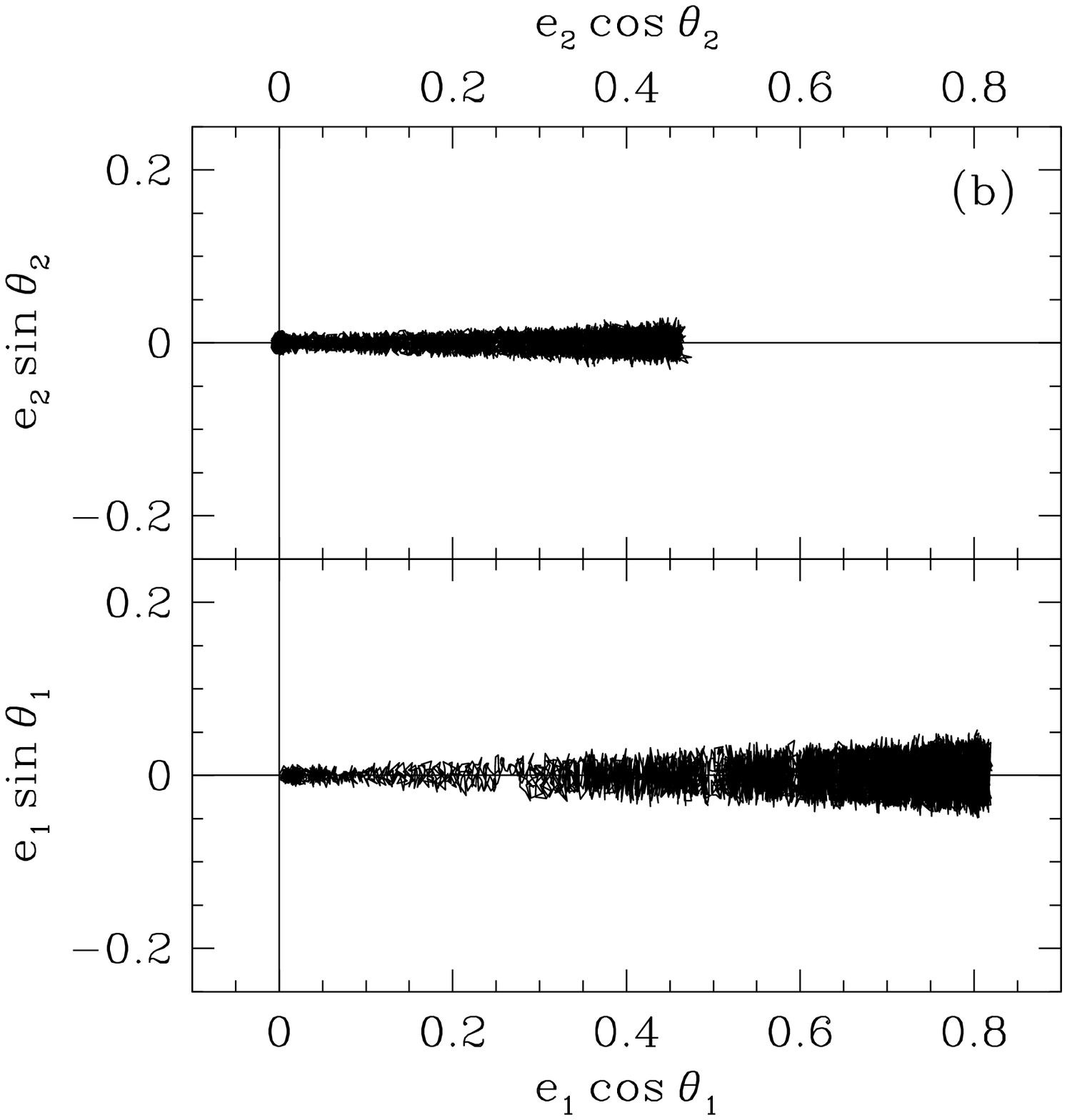}
\caption{
({\it a}) Time evolution of the semi-major axes and eccentricities and
({\it b}) evolution of the mean-motion resonance variables $\theta_1 =
\lambda_1 - 2\lambda_2 + \varpi_1$ and $\theta_2 = \lambda_1 - 2\lambda_2 +
\varpi_2$ in plots of $e_i \sin\theta_i$ versus $e_i \cos\theta_i$ for a
calculation where the outer planet is forced to migrate inward with
${\dot a}_2/a_2 = -5 \times 10^{-5}\,{\rm yr}^{-1}$ and there is no
eccentricity damping.
Both $\theta_1$ and $\theta_2$ are captured into small amplitude libration
about $0^\circ$, but the eccentricities exceed the observed values for the
\GJ\ planets ({\it dashed lines}) shortly after resonance capture, when
the semi-major axes of the resonantly locked planets decrease by only
$\approx 7\%$.
\label{figure3}}
\end{figure}

\begin{figure}[t]
\plotone{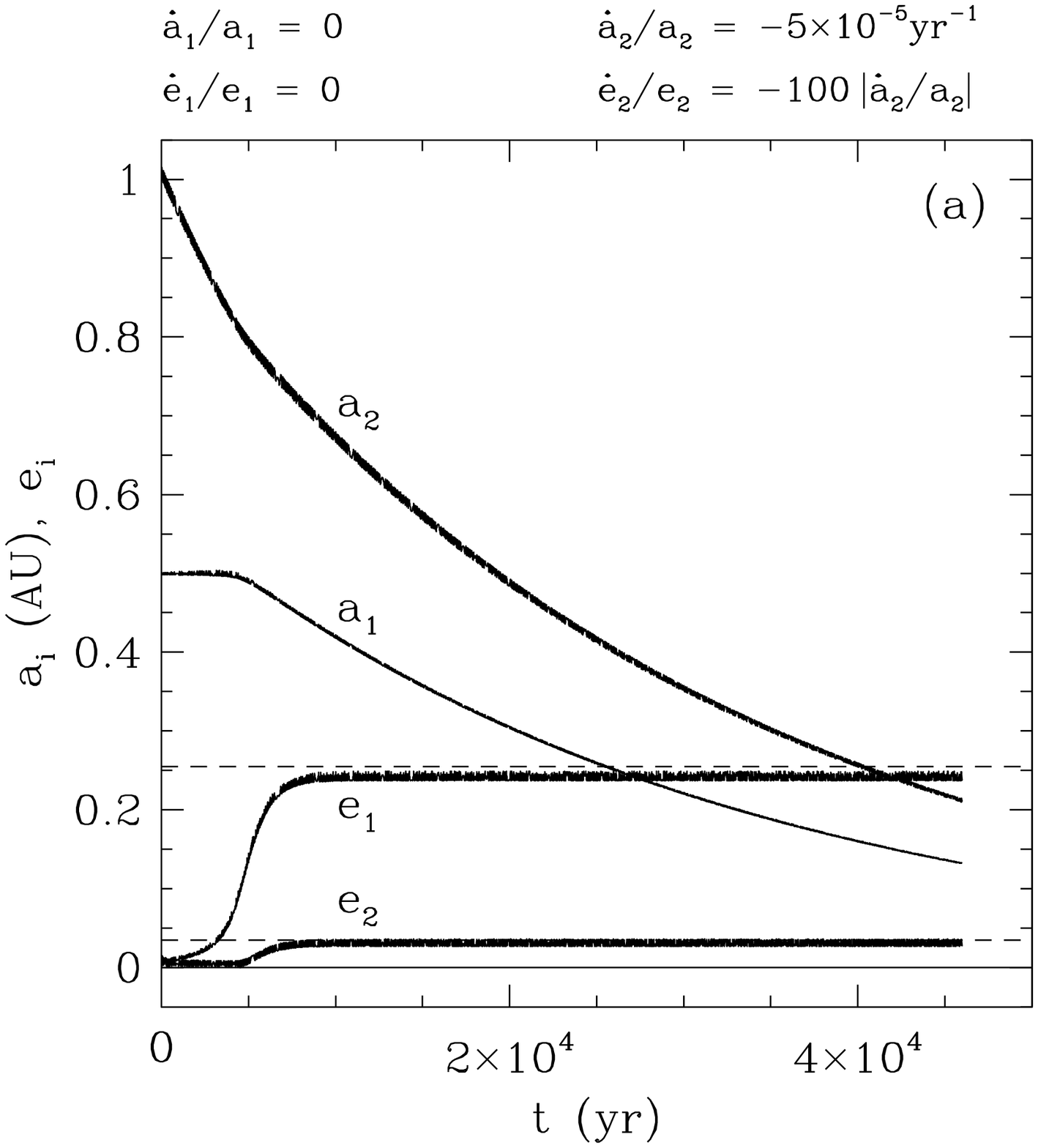}
\end{figure}
\begin{figure}[h]
\plotone{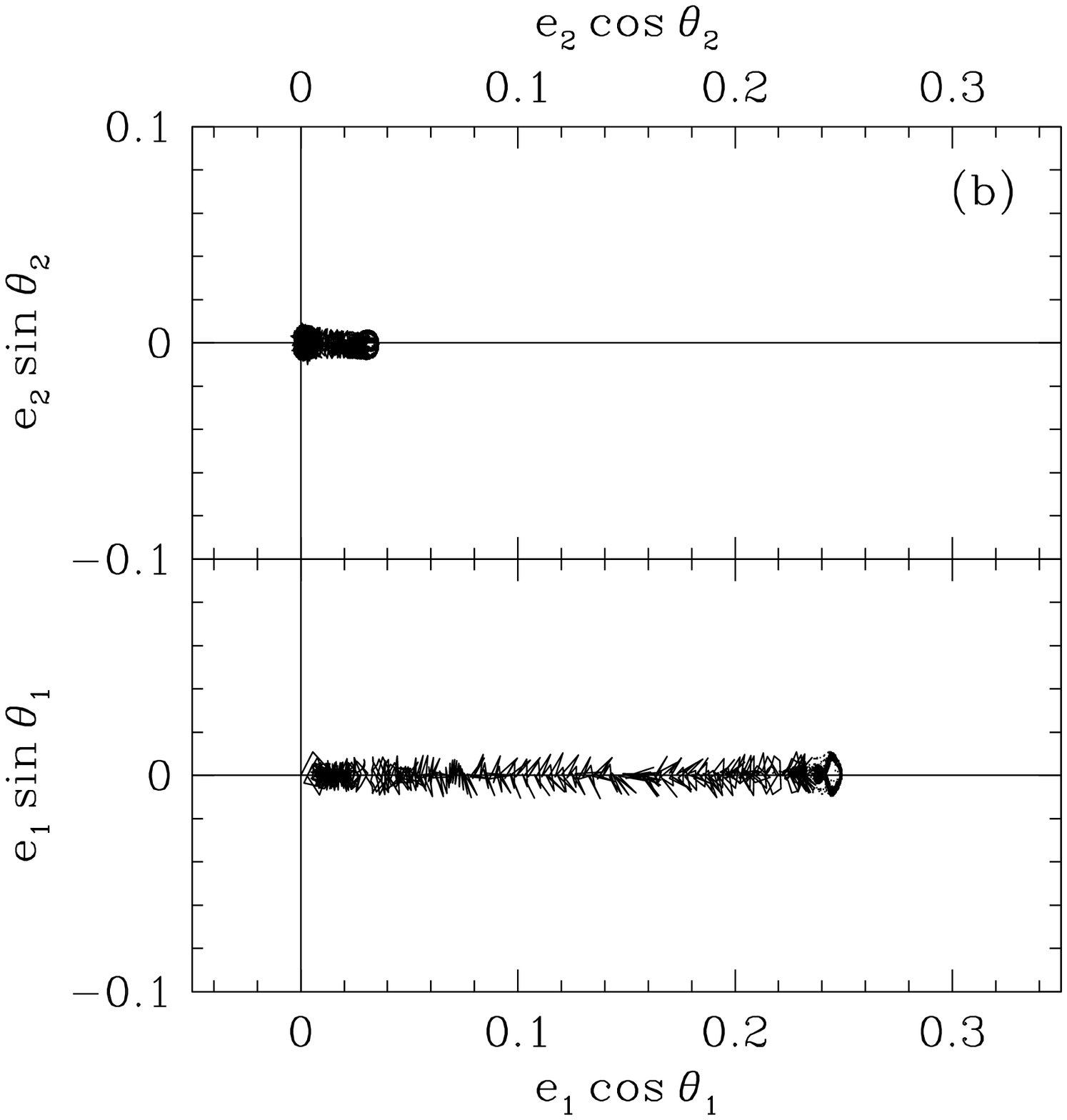}
\caption{
Same as Fig.~\ref{figure3}, but for a calculation with eccentricity damping
of the form ${\dot e}_2/e_2 = -K |{\dot a}_2/a_2|$, where $K = 100$.
After resonance capture, in addition to librations of $\theta_1$ and
$\theta_2$ about $0^\circ$, the eccentricities reach equilibrium values
close to the observed values for the \GJ\ planets ({\it dashed lines})
and remain constant for arbitrarily long migration in the resonances.
(The jagged nature of the plot ({\it b}) prior to equilibrium is due to
sparse sampling.)
\label{figure4}}
\end{figure}

The sequence of configurations with increasing eccentricities that the
system is driven through after resonance capture is in fact a sequence with
increasingly less negative periapse precession rates.
As we discussed in \S 2, the forced eccentricities (and the libration
centers) for a system with stable simultaneous librations of $\theta_1$
and $\theta_2$ are determined by the requirement that the longitudes of
periapse on average precess at the same rate (which in turn is determined
by ${\dot \lambda}_1 - 2 {\dot \lambda}_2 + {\dot \varpi}_i = 0$).
Although a longer integration indicates that the system does eventually
become unstable when $e_1 \approx 0.86$, the existence of stably librating
configurations with $e_1$ up to $0.86$ is remarkable.
In particular, the configurations with $e_1 \ga 0.71$ have {\it prograde}
periapse precessions.
We have integrated the configurations at $t = 2 \times 10^4 \yr$ (with
$e_1 \approx 0.59$ and retrograde periapse precessions) and $5 \times 10^4
\yr$ (with $e_1 \approx 0.81$ and prograde periapse precessions) forward
with migration turned off and confirmed that the system remains stable
as seemingly indefinitely repeating configurations.

As we can see in Figure~\ref{figure3}{\it a}, without eccentricity damping,
the eccentricities exceed the observed values for the \GJ\ planets (dashed
lines) when the semi-major axes of the resonantly locked planets have
decreased by only $\approx 7\%$ after capture into the resonances.
This result is insensitive to the adopted parameters.
If the migration rate is not too fast (see below), the evolution of a system
with different initial $a_2$ (but for convenience, the same $a_1/a_2$)
or ${\dot a}_2/a_2$ is
essentially identical to that shown in Figure~\ref{figure3} if we plot the
semi-major axes in units of initial $a_2$ and time in units of
$({\dot a}_2/a_2)^{-1}$.
Therefore, unless by coincidence resonance capture occurs just before
migration stops because of, e.g., nebula dispersal,
eccentricity damping is necessary to produce the observed eccentricities of
the \GJ\ system.

From a set of calculations similar to that shown in Figure~\ref{figure3},
with initial $a_1/a_2 = 1/2$, $a_2 = 1$, $2$, and $4\au$, and different
${\dot a}_2/a_2$, we find that certain capture of both 2:1 mean-motion
resonance variables requires
\begin{equation}
\left|{\dot a}_2 \over a_2\right| \la
3 \times 10^{-2} \left(a_2 \over {\rm AU}\right)^{-3/2} \yr^{-1} ,
\label{dadtlim}
\end{equation}
where $a_2$ is the semi-major axis of the outer planet when the 2:1
commensurability is encountered.
For migration rate within a factor of a few of, but below, the above limit,
both resonance variables are captured into libration, but the centers of
libration can be significantly different from $0^\circ$ and the amplitudes
of libration can be large.
The condition (\ref{dadtlim}) is satisfied by a migration rate due to
planet-nebular interaction with the nominal parameter values in equation
(\ref{vistime}) by almost 3 orders of magnitude.

To examine the effects of orbital eccentricities on capture into the 2:1
resonances,
we performed a series of calculations with non-zero initial eccentricities.
Four types of initial conditions were considered:
the outer planet was initially at either periapse or apoapse, and
either the initial $e_1$ or $e_2$ was fixed at $0.01$ and the other initial
eccentricity was varied.
The inner planet was always started at periapse, with its mean longitude
differing from that of the outer planet by $180^\circ$, and
the remaining parameters were identical to those used in the calculation
shown in Figure~\ref{figure3}.
Since gravitational interaction between the planets causes the
eccentricities to fluctuate even when $a_1$ and $a_2$ are close to their
initial values,
the eccentricities quoted below are the maximum eccentricities when $a_1$
and $a_2$ are close to their initial values and not the initial
eccentricities.
We find that certain capture of both 2:1 mean-motion resonance variables
requires
\begin{equation}
e_1 \la 0.06 \qquad {\rm and} \qquad e_2 \la 0.03.
\end{equation}
For eccentricities above these limits, there is non-zero probability for
the planets to be captured into higher order resonances (e.g., 5:2)
encountered before the 2:1 commensurability.

\subsection{Migration with Eccentricity Damping}

Figure~\ref{figure4} shows the results of a calculation similar to that
shown in Figure~\ref{figure3}, but with eccentricity damping of the form
${\dot e}_2/e_2 = -K |{\dot a}_2/a_2|$, where $K = 100$.
The capture of the resonance variables $\theta_1$ and $\theta_2$ into
libration about $0^\circ$ and the initial evolution after resonance capture
are similar to the case without damping.
However, the eccentricity growth eventually terminates when the damping
balances the excitation due to resonant interaction between the planets.
The eccentricities reach equilibrium values that remain constant for
arbitrarily long migration in the resonances.
With $K = 100$, the equilibrium eccentricities are close to the observed
eccentricities of the \GJ\ system.
At the end of the calculation shown in Figure~\ref{figure4}, when
$t = 4.6 \times 10^4 \yr$, the semi-major axes are also similar to those
of the \GJ\ planets.

\begin{figure}[h]
\plotone{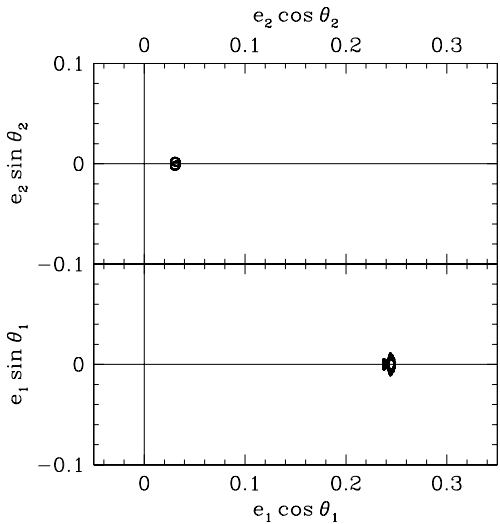}
\caption{
Continued small amplitude librations of the mean-motion resonance variables
$\theta_1$ and $\theta_2$ after termination of planet migration and
eccentricity damping.
The configuration at the end of the calculation shown in Fig.~\ref{figure4}
is integrated forward for $500\yr$ with migration and damping turned off.
\label{figure5}}
\end{figure}

\begin{figure}[h]
\plotone{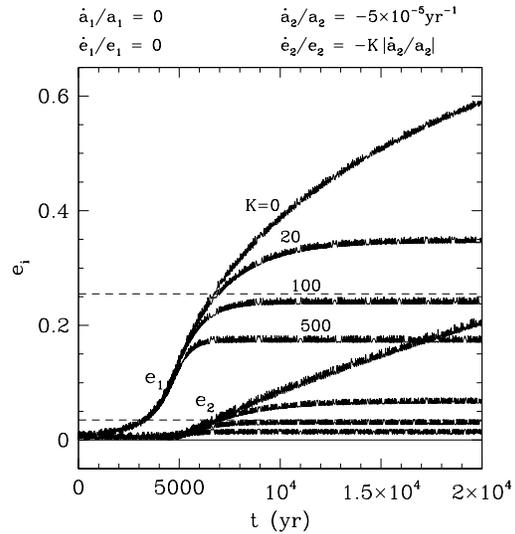}
\caption{
Decrease of the equilibrium eccentricities with increasing eccentricity
damping rate.
Time evolutions of the eccentricities for a set of calculations with
${\dot a}_2/a_2 = -5 \times 10^{-5}\,{\rm yr}^{-1}$,
${\dot e}_2/e_2 = -K |{\dot a}_2/a_2|$, and different $K$ are shown.
\label{figure6}}
\end{figure}

We have integrated the configuration at the end of the calculation shown
in Figure~\ref{figure4} forward with migration and damping turned off.
The system remains stable, with almost no change in the amplitudes of
libration (compare Figs.~\ref{figure4}{\it b} and \ref{figure5}), which
are somewhat smaller than those of the \GJ\ system (compare
Figs.~\ref{figure1} and \ref{figure5}).
Although the libration amplitudes are fairly similar in the two
Laughlin-Chambers best-fit solutions,
we may find in the future, as more data is acquired, that improved best-fit
solutions will have smaller libration amplitudes.
Alternatively, larger amplitudes could be generated after termination of
migration and damping by encounters with remaining planetesimals.

In Figure~\ref{figure6} we show the time evolutions of the eccentricities
for a set of calculations with ${\dot a}_2/a_2 = -5 \times
10^{-5}\yr^{-1}$, ${\dot e}_2/e_2 = -K |{\dot a}_2/a_2|$, and
different $K$.
The equilibrium eccentricities decrease with increasing $K$ and are
significantly different from the observed eccentricities of the \GJ\
system if $K$ is more than a factor 2--3 larger or smaller than
$K = 100$.

\begin{figure}[h]
\plotone{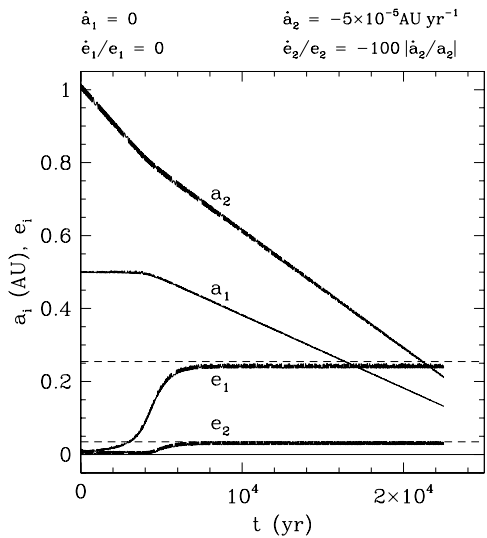}
\caption{
An example demonstrating that the equilibrium eccentricities are determined
by the ratio of damping to migration rates, $K$, and do not depend on the
functional form of the migration rate.
The results of a calculation with the same ratio ($K = 100$) as that shown
in Fig.~\ref{figure4} but with a different form of the migration rate
(${\dot a}_2 =$ constant) are shown.
\label{figure7}}
\end{figure}

As long as ${\dot e}_2/e_2 = -K |{\dot a}_2/a_2|$, where $K$ is a positive
constant, the equilibrium eccentricities are determined by the value of
$K$ only and are insensitive to either the magnitude or functional form of
${\dot a}_2/a_2$.
This is demonstrated in Figure~\ref{figure7}, where we plot the results
of a calculation with the same $K$($= 100$) as that shown in
Figure~\ref{figure4} but with a different form of the migration rate:
${\dot a}_2 =$ constant or ${\dot a}_2/a_2 \propto 1/a_2$.
The equilibrium eccentricities are identical in Figures~\ref{figure4} and
\ref{figure7}.
If ${\dot e}_2/e_2$ is not exactly proportional to $|{\dot a}_2/a_2|$, the
eccentricities would decrease slowly (after reaching maxima) or increase
slowly as the resonantly locked planets migrate, but the ratio of damping
to migration rates, $|({\dot e}_2/e_2)/({\dot a}_2/a_2)|$, just before
migration and damping stop should be close to $100$ for the final
eccentricities to be close to the observed values of the \GJ\ system.

\begin{figure}[h]
\plotone{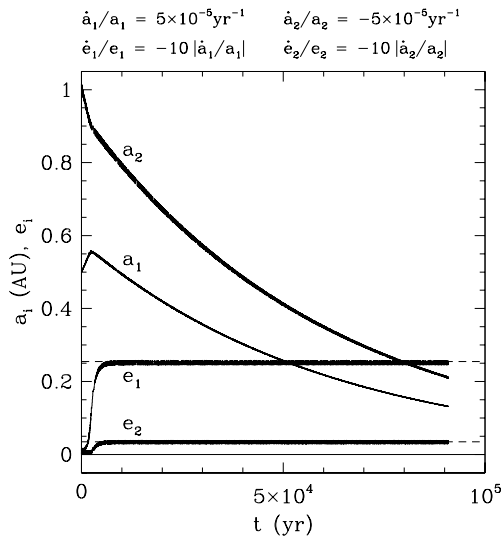}
\caption{
Time evolution of the semi-major axes and eccentricities for a calculation
where the inner planet is also forced to migrate outward, with
${\dot a}_1/a_1 = -{\dot a}_2/a_2 = 5 \times 10^{-5}\,{\rm yr}^{-1}$,
${\dot e}_i/e_i = -K |{\dot a}_i/a_i|$, and $K = 10$.
Even with eccentricity damping of both planets, $K \approx 10$ is required
to produce the observed eccentricities of the \GJ\ planets ({\it dashed
lines}).
\label{figure8}}
\end{figure}

Thus far we have considered forced inward migration of the outer planet
only.
This is the most likely situation in the planet-nebular interaction
scenario discussed in \S 3, since the inner disk is likely to be depleted
and the dominant interactions are expected to be those between the outer
disk and the outer planet.
However, if the inner disk is not depleted, at least initially, or
migration and damping are due to another process, the inner planet may
also be forced to migrate outward.
In Figure~\ref{figure8} we show the results of a calculation where, for
simplicity, the inner planet is forced to migrate outward at the same rate
that the outer planet is forced to migrate inward (${\dot a}_1/a_1 =
-{\dot a}_2/a_2 = 5 \times 10^{-5}\,{\rm yr}^{-1}$) and ${\dot e}_i/e_i =
-K |{\dot a}_i/a_i|$.
Initially, the inner (outer) planet migrates outward (inward) at the
prescribed rate.
After resonance capture, resonant interaction overcomes the forced outward
migration of the inner planet, and both planets migrate inward slowly.
The equilibrium eccentricities are close to the observed eccentricities of
the \GJ\ system when $K = 10$.
Therefore, even with eccentricity damping of both planets, significant
eccentricity damping with $|{\dot e}_i/e_i| \gg |{\dot a}_i/a_i|$ is
required to produce the observed orbital eccentricities of the \GJ\
planets.

\section{DISCUSSION}

\subsection{Eccentricity Damping by Tidal Dissipation in the Star and Planets}

We have found in \S 4 that significant eccentricity damping with
$|{\dot e}_i/e_i| \gg |{\dot a}_i/a_i|$ is required to produce the
observed eccentricities of the \GJ\ system if the migration has been at
all extensive after resonance capture.
As we discussed in \S 3, it is as yet unclear whether sufficient
damping could be produced by planet-nebula interaction, even if the
condition (\ref{2to1condII}) is satisfied and eccentricity damping is
expected (see also \S 5.2).
In this subsection we show that alternative eccentricity damping by tidal
dissipation within the star and planets during planet migration is
completely negligible.

The star \GJ\ would most likely still be in its pre-main-sequence
contracting phase during disk evolution and planet migration.
To estimate the rate of circularization of an orbit due to tidal dissipation
in the star,
we use the stellar radius when the time $t$ since initial contraction is
approximately the migration time scale of $|a/{\dot a}|=2\times 10^4\yr$
(eq.~[\ref{vistime}]) and keep the planets at their current distances
from the star.
The planets would have migrated for longer than the migration time scale
(see, e.g., Fig.~\ref{figure4}) and it would take time for the planets to
form.
But by adopting this minimum time to yield the maximum probable stellar
radius and by keeping the planets at their current distances,
we maximize our estimate of the tidal rate of circularization of an orbit.

Since the mass of \GJ\ is only $0.32 M_\odot$,
we assume that \GJ\ follows the nearly vertical Hayashi track in the HR
diagram and remains fully convective during its contracting phase.
Hence the star remains a polytrope of index $n=1.5$, with effective
temperature $T_e \approx 3500\,{\rm K}$ (e.g., D'Antona \& Mazzitelli 1994).
From the virial theorem, the stellar luminosity
\begin{eqnarray}
L = 4\pi R_0^2\sigma T_e^4
\!\!&=&\!\! \frac{1}{2}\frac{d}{dt}\left(\frac{3}{5-n}\frac{GM_0^2}{R_0}\right)
\nonumber \\ & & \nonumber \\
\!\!&=&\!\! \frac{-3}{2(5-n)}\frac{GM_0^2}{R_0^2}\frac{dR_0}{dt},
\label{eq:virial}
\end{eqnarray}
where $\sigma$ is the Stefan-Boltzmann constant, $G$ is the gravitational
constant, $M_0$ and $R_0$ are the stellar mass and radius, and the form of
the gravitational energy for a polytrope of index $n$ is given by
Chandrasekhar (1939).
Integration of equation (\ref{eq:virial}), with $3/2(5-n) = 3/7$ for a
polytrope of index $n=1.5$, gives $R_0 \approx 3.85\times 10^{11}\,{\rm cm}$
after $t = 2\times 10^4\yr$, leading to a ratio of the stellar radius to
the orbital semi-major axis $R_0/a\approx 0.197$ for the inner
planet and $R_0/a\approx 0.125$ for the outer planet.

Since the present orbital period of the inner planet of the \GJ\
system is about 30 days,
which is long compared to the periods of free oscillation of the star,
we can use the rate for circularizing an orbit corresponding to
dissipation dominated by the equilibrium tide for the star,
which is given by (Zahn 1989)
\begin{equation}
\frac{1}{t_{\rm circ}} = \left|\frac{\dot e}{e}\right| =
21 q (1+q) \frac{\lambda_{\rm circ}}{t_f} \left(\frac{R_0}{a}\right)^8 ,
\label{eq:tcirc}
\end{equation}
where $q = M/M_0$ is the planet-star mass ratio and $t_f =
(M_0 R_0^2/L)^{1/3}$.
We neglect the reduction of the turbulent viscosity when the tidal
period is short compared to the turnover time of the largest convective
eddies and adopt the maximum $\lambda_{\rm circ} \approx 0.048$ for a fully
convective star.
Substitution of $R_0/a$ from the previous paragraph, along with $q$ for
the Laughlin-Chambers Keck$+$Lick solution (Table~\ref{table1}),
$t_f = (M_0/4\pi \sigma T_e^4)^{1/3} \approx 0.574\yr$, and
$\lambda_{\rm circ} \approx 0.048$, into equation (\ref{eq:tcirc}) gives
\begin{eqnarray}
\frac{1}{t_{\rm circ}}
\!\!&\approx&\!\! 9.1\times 10^{-9}\yr^{-1} \quad {\rm ~for~inner~planet,}\nonumber\\
\!\!&\approx&\!\! 7.6\times 10^{-10}\yr^{-1}\quad {\rm for~outer~planet.}\nonumber\\
&&
\label{eq:tcirc2}
\end{eqnarray}
Since we have assumed that the planets are at their closest proximity to
the star for the whole time the tidal effects are damping the
eccentricities and that the star is inflated to a maximum size, these
values are extreme upper bounds on the rate of eccentricity damping by
tides raised on the star.
Even in its inflated state, the dissipation in the star can have
essentially no effect on the orbital eccentricities.

The theoretical circularization rate in equation (\ref{eq:tcirc}) has been
tested by comparing with the observed circularization rates of binaries.
It is in good agreement with the observed rates for binaries containing
giant stars (Verbunt \& Phinney 1995), but even with the maximum
$\lambda_{\rm circ}$, it is about $50$--$100$ times slower than the
observed rates for binaries containing main-sequence solar-type stars,
which have radiative cores (Claret \& Cunha 1997; Goodman \& Oh 1997).
It is not yet clear why there is a discrepancy in the latter case or that
this discrepancy is relevant for a fully convective pre-main-sequence star.
Nevertheless, even if we increase the circularization rates in equation
(\ref{eq:tcirc2}) by a factor $100$, they are still much smaller than
the migration rate due to planet-nebula interaction.

Tidal dissipation within a gaseous planet damps its orbital
eccentricity at a rate given by (e.g., Peale, Cassen, \& Reynolds 1980)
\begin{equation}
\left|\frac{\dot e}{e}\right|=\frac{21}{2}\frac{k_2}{q Q}
\frac{2\pi}{P} \left(\frac{R}{a}\right)^5,
\label{eq:planettide} 
\end{equation}
where $R$, $k_2$, and $Q$ are the radius, the potential Love number, and
the dissipation function of the planet.
If we adopt values for $R$, $k_2$, and $Q$ similar to those of Jupiter, with
$R = 7\times 10^4\,{\rm km}$, $k_2 = 0.38$ (Gavrilov \& Zharkov
1977), and $Q = 5\times 10^4$ (which is approximately the lower bound on
$Q$ for Jupiter; Yoder \& Peale 1981),
$|{\dot e}/e| \approx 1.5\times 10^{-12}\yr^{-1}$ for the inner planet of
the \GJ\ system.
The rate for damping the outer planet's orbital eccentricity from dissipation
within itself is of course even smaller.
The planets would most likely be larger and contracting during their
migration.
However, unless the planetary radii are unrealistically large and
comparable to the Roche radii, the eccentricity damping rate due to tidal
dissipation within the planets is smaller than the migration rate due to
planet-nebula interaction by several orders of magnitude.

\subsection{Other Studies}

After completing our calculations (Lee \& Peale 2001), two papers with
complementary calculations came to our attention
(Snellgrove, Papaloizou, \& Nelson 2001; Murray, Paskowitz, \& Holman 2001).

Snellgrove et al. (2001) also find from numerical orbit integrations with
forced migration and eccentricity damping of the outer planet that the
orbital eccentricities of the \GJ\ planets require a short time scale
for eccentricity damping compared to the migration time scale.
Note, however, that they adopt the minimum planetary masses from the
two-Kepler fit by Marcy et al. (2001) and, not surprisingly, have
difficulties matching both $e_1$ and $e_2$ from their calculations to those
from the same fit, since the eccentricities from the two-Kepler fit are
not consistent with small-amplitude simultaneous librations of $\theta_1$
and $\theta_2$.
Snellgrove et al. also develop an analytic resonance theory that is first
order in the eccentricities.
As we showed in \S 2, a first order theory is inadequate for understanding
the current \GJ\ resonance configuration, which has $\theta_1$,
$\theta_2$, and $\theta_3$ all librating about $0^\circ$.
Furthermore, as we found in \S 4 (see, e.g., Fig.~\ref{figure3}), because
the eccentricities are excited to large values in the \GJ\ evolution
before resonance capture, there is no time during the evolution when a
first order theory can be useful for this system.
Thus all predictions coming from this theory should be viewed with caution.
Snellgrove et al. also present a hydrodynamic simulation of the
planet-nebula interaction, where both planets are inside a cavity with
almost no disk material.
At the end of this simulation, $e_1 \approx 0.34$ and the semi-major axes
of the planets have decreased by about $13\%$ since resonance capture.
We have performed a numerical orbit integration {\it without} eccentricity
damping similar to that shown in Figure~\ref{figure3} but with the
planetary masses adopted by Snellgrove et al. and find that
$e_1 \approx 0.38$ for the same reduction in semi-major axes.
Thus it is not clear that the eccentricities have reached equilibrium values
at the end of this simulation;
even if they have, these equilibrium values are too large for the
\GJ\ system.
It appears that the disk model used in this hydrodynamic simulation is
not very effective in damping the eccentricities, which emphasizes the
uncertainty in such damping.

Murray et al. (2001) consider mainly an alternative scenario in which the
migration and eccentricity damping of the outer planet are due to
scattering of planetesimals in the disk population.
Their numerical simulations confirm our results (and those of Snellgrove
et al. 2001) of easy capture of an inner planet into resonance, the
dual migration of both planets in the resonance, and the growth of the
eccentricities, although they limit their numerical and analytical studies
to an outer planet of Jupiter mass and inner planets of several Earth masses
and devote much of the discussion to resonances of higher order than those
at the 2:1 mean-motion commensurability.
Thus their results are not directly applicable to the \GJ\ system.

\section{CONCLUSIONS}

Radial velocity measurements by Marcy et al. (2001) have revealed two
planets in resonant orbits about the star \GJ.
The remarkable orbital fit obtained by Laughlin \& Chambers (2001),
which finds both lowest order, eccentricity-type mean-motion resonance
variables at the 2:1 commensurability librating with small amplitudes,
means that the resonances are almost certainly real and indefinitely
stable.
The existence of the eccentricity-type resonances implies that the assumed
coplanarity of the orbits is probably close to reality.

The \GJ\ planetary system has revealed properties of the 2:1 orbital
resonances that have not been observed nor analyzed before.
The libration of both lowest order mean-motion resonance variables,
$\theta_1 = \lambda_1 - 2 \lambda_2 + \varpi_1$ and
$\theta_2 = \lambda_1 - 2 \lambda_2 + \varpi_2$,
and the secular resonance variable, $\theta_3 = \varpi_1 - \varpi_2$, about
$0^\circ$ was not anticipated, as the 
familiar Io-Europa 2:1 resonance has $\theta_1$ librating about
$0^\circ$, but $\theta_2$ and $\theta_3$ librating about $180^\circ$ ---
a configuration that would persist in the absence of Ganymede.
Thus conjunctions for the Jovian satellites occur when Io is near
periapse and Europa is near apoapse, whereas conjunctions of the
two planets about \GJ\ occur when both planets are near periapse.
We understood this to be mainly a function of the eccentricities of the
orbits, where the resonance configuration with $\theta_1 \approx 0^\circ$
and $\theta_2\approx\theta_3\approx 180^\circ$ must obtain when the
eccentricities are small, but the resonance configuration with
$\theta_1\approx\theta_2\approx\theta_3\approx0^\circ$ prevails for a
system with masses like those in \GJ\ when the eccentricities are large.
A necessary condition for stable simultaneous librations of both
mean-motion resonance variables is that $\dot\varpi_1 = \dot\varpi_2$ on
average, so that the relative alignment of the lines of apsides of the two
orbits is maintained.
The periapse precessions are dominated by resonant terms in the disturbing
potential whose arguments are $\theta_1$, $\theta_2$, $\theta_3$, and
their linear combinations.
The dominance of the lowest order terms when the eccentricities are small
allows equal precession rates only if the lines of apsides are anti-aligned
($\theta_1\approx 0^\circ$, $\theta_2\approx\theta_3\approx 180^\circ$),
whereas dominance of higher order terms when the eccentricities are large
results in equal precession rates for a system with masses like those in
\GJ\ when $\theta_1\approx\theta_2\approx\theta_3\approx0^\circ$.
The equality of the precession rates also determines a relationship between
the eccentricities of the two orbits, although there is no simple analytic
expression for this relationship when the eccentricities are large,
since the first order theory is not a good representation.
The stability of the \GJ\ resonance configuration for values of $e_1$
up to 0.86 was also a surprise.

Any process that drives the two originally widely separated orbits
toward each other can result in capture of the planets into orbital resonances
at the 2:1 commensurability.
The naturally occurring situation where nebular disk material is cleared
between two planets sufficiently massive to individually open gaps in the
disk (Bryden et al. 2000; Kley 2000) leads to inward migration of the
outer planet and possibly outward migration of the inner planet.
Thus the likely origin of the resonances in the \GJ\ system is this
differential planet migration due to torques induced by the planet-nebula
interaction.
We have shown that forced inward migration of the outer planet of the \GJ\
system results in certain capture of $\theta_1$, $\theta_2$ and hence
$\theta_3$ into libration if initially $e_1\la 0.06$ and $e_2\la 0.03$ and
$|\dot a_2/a_2|\la 3\times 10^{-2}(a_2/\au)^{-3/2}\yr^{-1}$.
The latter rate is three orders of magnitude higher than the likely
rate of $\sim 5\times 10^{-5} (a_2/\au)^{-3/2}\yr^{-1}$ due to
planet-nebular interaction.
The bounds on the eccentricities result not so much from the transition
from certain to probabilistic capture at the 2:1 resonances but from likely
capture into higher order resonances such as 5:2 before the 2:1
commensurability is encountered. 

Continued migration of the planets while locked in the 2:1 resonances
leads to rapid growth in the orbital eccentricities that exceed the
observed eccentricities of the \GJ\ system after only a further decrease
in the semi-major axes of about 7\% if there is no eccentricity damping.
So unless resonance capture occurred near the end of migration,
the observed values of the eccentricities require eccentricity damping.
With damping of the form ${\dot e}_i/e_i=-K|{\dot a}_i/a_i|$, where $K$ is
a positive constant,
eccentricity growth is terminated at values of the eccentricities that
increase with decreasing $K$, and the eccentricities remain constant for
indefinite duration of the migration.
The observed eccentricities result for $K\approx 100$ if there is
forced migration and eccentricity damping of the outer planet only,
but for $K\approx 10$ if there is also forced migration and eccentricity
damping of the inner planet.
This result is independent of the magnitude or functional form of
${\dot a}_i/a_i$ as long as ${\dot e}_i/e_i=-K|{\dot a}_i/a_i|$ is preserved.
Relaxing the last condition leads to a slow drift in the eccentricities
during migration and would require the migration to terminate as the
eccentricities pass through the observed values.

Existing analytic estimates of the effects of planet-nebular interaction
are consistent with eccentricity damping of the form
${\dot e}_i/e_i=-K|{\dot a}_i/a_i|$, if the planet-star mass ratio is not
too large (e.g., Goldreich \& Tremaine 1980; Artymowicz 1992, 1993;
Lin \& Papaloizou 1993; Ward 1997).
However, the planet-star mass ratio of the outer planet of the \GJ\
system is sufficiently close to the critical value separating eccentricity
growth from damping for nominal values of the disk parameters that it is
uncertain whether such damping would occur.
Even if the disk parameters are such that eccentricity damping would occur,
it is not clear that the magnitude of $K$ would be sufficiently large to
constrain the eccentricities in the \GJ\ system to the observed values.
We have shown that the alternative eccentricity damping mechanism involving
the dissipation of tidal energy within the star and the planets is
completely negligible.
Further long-term hydrodynamic simulations with different physical
assumptions and parameters are required to determine whether
planet-nebular interaction could produce sufficient eccentricity damping to
allow arbitrary migration of the planets within the resonances in the
young \GJ\ system while preserving eccentricities comparable to those
observed.
If not, the migration must have been finely tuned to stop when the system
had progressed to its observed state, although this latter constraint is
too ad hoc to be believable.

\acknowledgments
It is a pleasure to thank Greg Laughlin for furnishing the details of the
dynamical fit by Laughlin \& Chambers before publication and Lars Bildsten
for pointing out the enhancement of tidal dissipation in a pre-main-sequence
star.
We also thank D.N.C. Lin, J.J. Lissauer, and W.R. Ward for informative
discussions.
This research was supported in part by NASA grants PGG NAG5-3646 and
OSS NAG5-7177.

\clearpage

\appendix

\section{NUMERICAL METHODS}

In this appendix we describe how the symplectic integrator SyMBA and the
Bulirsch-Stoer integrator used for the numerical orbit integrations
presented in \S 4 were modified to include the forced orbital migration
and eccentricity damping terms.

A second-order symplectic integrator for a Hamiltonian of the
form $H = H_0 + H_1$, where $H_0$ and $H_1$ are separately integrable,
can be represented as
\begin{equation}
E_0\left(\tau \over 2\right)
E_1\left(\tau\right)
E_0\left(\tau \over 2\right) .
\label{SIA2}
\end{equation}
The three operators in equation (\ref{SIA2}) represent a single step of an
algorithm that starts with evolving the system under the influence of $H_0$
only for half a time step $\tau/2$, then evolving it for a full time step
$\tau$ under the influence of $H_1$, and then evolving it for another half
a time step $\tau/2$ under $H_0$.
For example, in the Wisdom-Holman (1991) method for the gravitational
$N$-body problem, the Hamiltonian in Jacobi coordinates is divided into $H_0$
that describes the Keplerian motion of the planets around a central star
and $H_1$ that describes the perturbation of the planets on one another and
is a function of the positions $\mbox{\boldmath $q$}_i$ only.
A step of the Wisdom-Holman method is thus:
(1) Each planet evolves along a Kepler orbit for time $\tau/2$;
(2) each planet receives a kick to its momentum of the amount
$-\tau \partial H_1/\partial \mbox{\boldmath $q$}_i$ while
$\mbox{\boldmath $q$}_i$ is unchanged (since $H_1$ does not involve the
canonical momenta);
(3) each planet evolves along a Kepler orbit for time $\tau/2$, starting
with the new momentum after the kick.
Recursive application of the basic algorithm (\ref{SIA2}) allows one to
construct symplectic integrators for Hamiltonians that consist of more than
two integrable parts.
For example, a single step of an algorithm for a Hamiltonian of the form
$H = H_0 + H_1 + H_2$ is
$E_0(\tau/2) E_1(\tau/2) E_2(\tau) E_1(\tau/2) E_0(\tau/2)$.

The symplectic integrator SyMBA (Duncan et al. 1998) is based on a variant
of the Wisdom-Holman method,
with the gravitational $N$-body Hamiltonian written in terms of positions
relative to a central star and barycentric momenta,
and employs a multiple time step technique to handle close encounters.
In the SyMBA algorithm, the Hamiltonian is divided into more than two parts
and the recursive application of the algorithm (\ref{SIA2}) discussed in
the previous paragraph is utilized.
Although the additional forced orbital migration and eccentricity damping
terms are not Hamiltonian, they can be incorporated in an analogous manner.
Our modified algorithm is
\begin{equation}
E_a\left(\tau \over 2\right)
E_e\left(\tau \over 2\right)
E_{\rm grav}\left(\tau\right)
E_e\left(\tau \over 2\right)
E_a\left(\tau \over 2\right) ,
\end{equation}
where $E_{\rm grav}(\tau)$ denotes a complete time step for the conservative
gravitational $N$-body problem using the SyMBA algorithm, and
$E_a(\tau/2)$ and $E_e(\tau/2)$ denote changing the canonical
variables according to the imposed ${\dot a}_i$ and ${\dot e}_i$ terms,
respectively, for time $\tau/2$.
During the application of the ${\dot a}_i$ term, all of the other orbital
elements are constant, and $a_{i,1} = a_{i,0} \exp(\tau {\dot a}_i/2 a_i)$,
where $a_{i,0}$ and $a_{i,1}$ are $a_i$ at the beginning and end of the
step, if ${\dot a}_i/a_i =$ constant (this can be easily generalized for,
e.g., ${\dot a}_i/a_i \propto a_i^\gamma$).
Note that we do not use truncated approximation such as $a_{i,1} = a_{i,0} 
(1 + \tau {\dot a}_i/2 a_i)$.
Similarly, the $E_e(\tau/2)$ step changes the eccentricities according to
$e_{i,1} = e_{i,0} \exp(-\tau K |{\dot a}_i/a_i|/2)$ if ${\dot e}_i/e_i =
-K|{\dot a}_i/a_i|$.
By modifying the algorithm in a symmetric manner and using exact solutions
in the $E_a(\tau/2)$ and $E_e(\tau/2)$ parts, there should be little (if
any) secular growth in the energy error (e.g., Mikkola 1998).

For the Bulirsch-Stoer integrator, additional terms must be included
in the equations of motion in Cartesian coordinates to account
for the forced orbital migration and eccentricity damping.
In the following, we simplify the notation by considering a specific planet
and dropping the subscript.
Let $(x,y,z)$ be the Cartesian coordinates of the planet with respect to
the star and $r$ be the distance of the planet from the star.
The osculating orbital elements $a$, $e$, $i$, $f$, $\omega$, and $\Omega$
are the semi-major axis, eccentricity, inclination, true anomaly, argument
of periapse, and longitude of the ascending node on the $xy$ plane,
respectively.
The additional terms in the equations of motion due to the forced migration
${\dot a}$ and eccentricity damping ${\dot e}$ terms are
\begin{eqnarray}
\frac{dx}{dt}\Bigg|_{\dot a} + \frac{dx}{dt}\Bigg|_{\dot e}
&=& \frac{\partial x}{\partial a}{\dot a} +
    \frac{\partial x}{\partial e}{\dot e} ,
\label{eq:dxdt}\\
& & \nonumber\\
\frac{d\dot x}{dt}\Bigg|_{\dot a} + \frac{d\dot x}{dt}\Bigg|_{\dot e}
&=& \frac{\partial\dot x}{\partial a}{\dot a} +
    \frac{\partial\dot x}{\partial e}{\dot e} ,
\label{eq:ddotxdt}
\end{eqnarray}
with similar expressions for the other coordinates.
To evaluate the partial derivatives in equations (\ref{eq:dxdt}) and
(\ref{eq:ddotxdt}) and similar expressions for the other coordinates,
we need to express the position and velocity in terms of the osculating
orbital elements:
\begin{eqnarray}
x &=& r\cos{\Omega}\cos{(\omega+f)}-r\cos{i}\sin{\Omega}\sin{(\omega+f)},
\nonumber\\
y &=& r\sin{\Omega}\cos{(\omega+f)}+r\cos{i}\cos{\Omega}\sin{(\omega+f)},
\label{eq:x}\\
z &=& r\sin{i}\sin{(\omega+f)},\nonumber
\end{eqnarray}
and
\begin{eqnarray}
{\dot x} &=& \cos{\Omega}[{\dot r}\cos{(\omega+f)}
                          -r{\dot f}\sin{(\omega+f)}] \nonumber\\
&& \quad    -\cos{i}\sin{\Omega}[{\dot r}\sin{(\omega+f)}
                                 +r{\dot f}\cos{(\omega+f)}], \nonumber\\
{\dot y} &=& \sin{\Omega}[{\dot r}\cos{(\omega+f)}
                          -r{\dot f}\sin{(\omega+f)}] \label{eq:dotx}\\
&& \quad    +\cos{i}\cos{\Omega}[{\dot r}\sin{(\omega+f)}
                                 +r{\dot f}\cos{(\omega+f)}], \nonumber\\
{\dot z} &=& \sin{i}[{\dot r}\sin{(\omega+f)}
                     +r{\dot f}\cos{(\omega+f)}], \nonumber
\end{eqnarray}
where $r$, ${\dot r}$, and $r{\dot f}$ are in terms of $a$, $e$, and $f$
(e.g., Murray \& Dermott 1999).
We also need
\begin{eqnarray}
\frac{\partial r}{\partial a} &=& \frac{r}{a},\nonumber \\
& & \nonumber \\
\frac{\partial r}{\partial e} &=& \left[-\frac{2er}{1-e^2}-\frac{r^2\cos{f}}
                                        {a(1-e^2)}\right],\nonumber\\
& & \nonumber \\
\frac{\partial\dot r}{\partial a} &=& -\frac{\dot r}{2a},\nonumber\\
& & \nonumber \\
\frac{\partial\dot r}{\partial e} &=& \frac{\dot r}{e(1-e^2)},
                                                 \label{eq:drdtae}\\
& & \nonumber \\
\frac{\partial (r\dot f)}{\partial a} &=& -\frac{r\dot f}{2a},\nonumber\\
& & \nonumber \\
\frac{\partial (r\dot f)}{\partial e} &=& \frac{r\dot f(e+\cos{f})}
                                          {(1-e^2)(1+e\cos{f})}.\nonumber
\end{eqnarray}
From equations (\ref{eq:dxdt}), (\ref{eq:x}), and (\ref{eq:drdtae}),
we find that
\begin{equation}
\frac{dx}{dt}\Bigg|_{\dot a} + \frac{dx}{dt}\Bigg|_{\dot e}
= \frac{x}{a}{\dot a}
+\left[\frac{r}{a(1-e^2)}-\frac{1+e^2}{1-e^2}\right]\frac{x}{e}{\dot e};
\label{eq:dxdtf}
\end{equation}
the additional terms for $dy/dt$ and $dz/dt$ are similar.
The additional terms for each of $d\dot x/dt$, $d\dot y/dt$, $d\dot z/dt$ are
distinct for variations in $e$, and we have
\begin{eqnarray}
\frac{d\dot x}{dt}\Bigg|_{\dot a} + \frac{d\dot x}{dt}\Bigg|_{\dot e}
&=& -\frac{\dot x}{2a} {\dot a} + \cos{\Omega}\left[
\frac{\partial \dot r}{\partial e}\cos{(\omega+f)}-\frac{\partial (r\dot
f)}{\partial e}\sin{(\omega+f)}\right]{\dot e}\nonumber\\
&&\nonumber\\
&&\quad -\cos{i}\sin{\Omega}\left[\frac{\partial\dot r}{\partial e}
\sin{(\omega+f)}+\frac{\partial(r\dot f)}{\partial e}\cos{(\omega+f)}
\right]{\dot e},\nonumber\\
&&\nonumber\\
\frac{d\dot y}{dt}\Bigg|_{\dot a} + \frac{d\dot y}{dt}\Bigg|_{\dot e}
&=& -\frac{\dot y}{2a} {\dot a}+\sin{\Omega}\left[
\frac{\partial \dot r}{\partial e}\cos{(\omega+f)}-\frac{\partial (r\dot
f)}{\partial e}\sin{(\omega+f)}\right]{\dot e}\label{eq:ddotxdtf}\\
&&\nonumber\\
&&\quad +\cos{i}\cos{\Omega}\left[\frac{\partial \dot r}{\partial e}
\sin{(\omega+f)}+\frac{\partial (r\dot f)}{\partial e}\cos{(\omega+f)}
\right]{\dot e},\nonumber\\  
&&\nonumber\\
\frac{d\dot z}{dt}\Bigg|_{\dot a} + \frac{d\dot z}{dt}\Bigg|_{\dot e}
&=& -\frac{\dot z}{2a} {\dot a}+\sin{i}\left[
\frac{\partial \dot r}{\partial e}\sin{(\omega+f)}+\frac{\partial (r\dot f)}
{\partial e}\cos{(\omega+f)}\right]{\dot e},\nonumber 
\end{eqnarray}
where equation (\ref{eq:drdtae}) should be used to get the functional forms.
Unfortunately, the orbital elements must be calculated at each call to
the differential equations when there is eccentricity damping.

\clearpage

\begin{deluxetable}{ccccccc}
\tablecolumns{7}
\tablewidth{0pt}
\tablecaption{Best Fit Orbital Parameters for the \GJ\ Planets
\label{table1}}
\tablehead{
\colhead{} & \multicolumn{2}{c}{2-Kepler Fit\tablenotemark{a}} &
\multicolumn{4}{c}{Dynamical Fit\tablenotemark{a}} \\
\colhead{} & \multicolumn{2}{c}{Keck+Lick} & \multicolumn{2}{c}{Keck} &
\multicolumn{2}{c}{Keck+Lick} \\
\colhead{} & \colhead{} & \colhead{} & \multicolumn{2}{c}{$\sin{i}=0.55$} &
\multicolumn{2}{c}{$\sin{i}=0.78$} \\
\colhead{Parameter\tablenotemark{b}} & \colhead{Inner} & \colhead{Outer} &
\colhead{Inner} & \colhead{Outer} & \colhead{Inner} &\colhead{Outer}
}
\startdata
$M$ ($M_{\s J}$)&$0.56/\sin{i}$&$1.89/\sin{i}$&$1.06$&$3.39$&$0.766$&$2.403$\\
$P$ (day)&30.12&61.02&29.995&62.092&30.569&60.128\\
$a$ (AU)&0.130&0.208&0.1294&0.2108&0.1309&0.2061\\
$e$&0.27&0.10&0.314&0.051&0.244&0.039\\
$\varpi$ (deg)&330&333&51.8&40.0&159.1&163.3\\
$T$ (JD)&\multicolumn{2}{c}{2450091.6}&\multicolumn{2}{c}{2450602.09}&\multicolumn{2}{c}{2449679.63}\\
${\cal M}$ (deg)&0.0&$-85.9$&289&340&356&173\\
\enddata
\tablenotetext{a}{Two-Kepler fit by Marcy et al. 2001 and dynamical fit
using a Levenberg-Marquardt $N$-body integration scheme by Laughlin \&
Chambers 2001.}
\tablenotetext{b}{The parameters are the planetary mass $M$ in terms of
Jupiter mass
$M_{\s J}$, the period $P$, the semi-major axis $a$, the orbital eccentricity
$e$, the longitude of the periapse $\varpi$, the epoch $T$, and the mean
anomaly ${\cal M}$. The stellar mass is $0.32M_\odot$.}
\end{deluxetable}

\clearpage

\begin{deluxetable}{lccccccc}
\tablecolumns{8}
\tablewidth{0pt}
\tablecaption{Comparison of contribution of various terms to the
periapse precession rates for the Io-Europa and \GJ\ systems
\label{table2}}
\tablehead{
\colhead{} & \colhead{} & \colhead{} &
\multicolumn{2}{c}{Io-Europa} & \colhead{} &
\multicolumn{2}{c}{\GJ\ Planets} \\
\cline{4-5} \cline{7-8} \\[-6pt]
\colhead{} & \colhead{} & \colhead{} &
\multicolumn{2}{c}{$\theta_1=0^\circ$, $\theta_2=\theta_3=180^\circ$} & \colhead{} &
\multicolumn{2}{c}{$\theta_1=\theta_2=\theta_3=0^\circ$} \\
\colhead{} & \colhead{} & \colhead{} &
\multicolumn{2}{c}{$e_1=0.0026$, $e_2=0.0013$} & \colhead{} &
\multicolumn{2}{c}{$e_1=0.255$, $e_2=0.035$} \\
\cline{4-5} \cline{7-8} \\[-6pt]
\colhead{} & \colhead{} & \colhead{} &
\colhead{$d\varpi_1/dt$} & \colhead{$d\varpi_2/dt$} & \colhead{} &
\colhead{$d\varpi_1/dt$} & \colhead{$d\varpi_2/dt$} \\
Terms & Order\tablenotemark{a} & \colhead{} &
\colhead{($\degperday$)} & \colhead{($\degperday$)} & \colhead{} &
\colhead{($\degperday$)} & \colhead{($\degperday$)}
}
\startdata
$\cos{k\theta_3}$ & $e^2$ & & \phs0.0034 & \phs0.0092 & & \phs0.0374 & $-0.0470$\\
& $e^4$ & & \phs0.0000 & \phs0.0000 & & \phs0.0032 & $-0.0080$\\[6pt]
$\cos{(k \theta_1 - \ell \theta_2)}$ & $e^1$ & & $-1.4832$ & $-1.5778$ & & $-0.2503$ & \phs0.1677\\
& $e^2$ & & \phs0.0190 & \phs0.0820 & & \phs0.1456 & $-0.3983$\\ 
& $e^3$ & & $-0.0002$ & $-0.0011$ & & $-0.0809$ & \phs0.2471\\
\cline{1-8} \\[-6pt]
Total & & & $-1.46$\phn\phn & $-1.49$\phn\phn & & $-0.145$\phn & $-0.039$\phn\\
Actual& & & & & & \multicolumn{2}{c}{$-0.116$} \\
\enddata
\tablenotetext{a}{Terms of order $e_1^{|k| + 2m} e_2^{|\ell| + 2n}$, with
$|k| + |\ell| + 2m + 2n = N$, in the disturbing potential (eq.~[\ref{phiexp}])
are grouped together under order $e^N$.}
\end{deluxetable}


\begin{thebibliography}{}
\bibitem[Artymowicz(1992)]{art92}
     Artymowicz, P. 1992, \pasp, 104, 769
\bibitem[Artymowicz(1993)]{art93}
     Artymowicz, P. 1993, \apj, 419, 166
\bibitem[Brouwer \& Clemence(1961)]{bro61}
     Brouwer, D., \& Clemence, G. M. 1961, Methods of Celestial Mechanics
     (New York: Academic)
\bibitem[Bryden et al.(2000)]{bry00}
     Bryden, G., R\'o\.zyczka, M., Lin, D. N. C., \& Bodenheimer, P. 2000,
     \apj, 540, 1091
\bibitem[Chandrasekhar(1939)]{cha39}
     Chandrasekhar, S. 1939, An Introduction to the Study of Stellar
     Structure (Chicago: Univ. of Chicago Press)
\bibitem[Claret \& Cunha(1997)]{cla97}
     Claret, A., \& Cunha, N. C. S. 1997, \aap, 318, 187
\bibitem[D'Antona \& Mazzitelli(1994)]{dan94}
     D'Antona, F., \& Mazzitelli, I. 1994, \apjs, 90, 467
\bibitem[Duncan et al.(1998)]{dun98}
     Duncan, M. J., Levison, H. F., \& Lee, M. H. 1998, \aj, 116, 2067
\bibitem[Gammie(1996)]{gam96}
     Gammie, C. F. 1996, \apj, 457, 355
\bibitem[Gavrilov \& Zharkov(1977)]{gav77}
     Gavrilov, S. V., \& Zharkov, V. N. 1977, Icarus, 32, 443
\bibitem[Goldreich \& Tremaine(1980)]{gol80}
     Goldreich, P., \& Tremaine, S. 1980, \apj, 241, 425
\bibitem[Goodman \& Oh(1997)]{goo97}
     Goodman, J., \& Oh, S. P. 1997, \apj, 486, 403
\bibitem[Goodman \& Rafikov(2001)]{goo01}
     Goodman, J., \& Rafikov, R. R. 2001, \apj, 552, 793
\bibitem[Hartmann et al.(1998)]{har98}
     Hartmann, L., Calvet, N., Gullbring, E., \& D'Alessio, P. 1998,
     \apj, 495, 385
\bibitem[Kley(2000)]{kle00}
     Kley, W. 2000, \mnras, 313, L47
\bibitem[Laughlin \& Chambers(2001)]{lau01}
     Laughlin, G., \& Chambers, J. E. 2001, \apjl, 551, L109
\bibitem[Lee \& Peale(2001)]{lee01}
     Lee, M. H., \& Peale, S. J. 2001, \baas, 33, in press
\bibitem[Lin \& Papaloizou(1993)]{lin93}
     Lin, D. N. C., \& Papaloizou, J. C. B. 1993, in Protostars and Planets
     III, ed. E. H. Levy \& J. I. Lunine (Tucson: Univ. Arizona Press), 749
\bibitem[Marcy et al.(2001)]{mar01}
     Marcy, G. W., Butler, R. P., Fischer, D., Vogt, S. S., Lissauer, J. J.,
     \& Rivera, E. J. 2001, \apj, 556, 296
\bibitem[Mikkola(1998)]{mik98}
     Mikkola, S. 1998, Celest. Mech. Dyn. Astron., 68, 249
\bibitem[Murray \& Dermott(1999)]{mur99}
     Murray, C. D., \& Dermott, S. F. 1999, Solar System Dynamics
     (Cambridge: Cambridge Univ. Press)
\bibitem[Murray et al.(2001)]{mur01}
     Murray, N., Paskowitz, M., \& Holman, M. 2001, preprint
     (astro-ph/0104475)
\bibitem[Papaloizou et al.(2001)]{pap01}
     Papaloizou, J. C. B., Nelson, R. P., \& Masset, F. 2001, \aap, 366, 263
\bibitem[Peale(1986)]{pea86}
     Peale, S. J. 1986, in Satellites, ed. J. A. Burns \& M. S. Matthews
     (Tucson, Univ. Arizona Press), 159
\bibitem[Peale(1999)]{pea99}
     Peale, S. J. 1999, \araa, 37, 533
\bibitem[Peale et al.(1980)]{pea80}
     Peale, S. J., Cassen, P., \& Reynolds, R. T. 1980, Icarus, 43, 65
\bibitem[Rivera \& Lissauer(2001)]{riv01}
     Rivera, E. J., \& Lissauer, J. J. 2001, \apj, 558, 392
\bibitem[Snellgrove et al.(2001)]{sne01}
     Snellgrove, M. D., Papaloizou, J. C. B., \& Nelson, R. P. 2001, \aap,
     374, 1092
\bibitem[Stone et al.(2000)]{sto00}
     Stone, J. M., Gammie, C. F., Balbus, S. A., \& Hawley, J. F. 2000,
     in Protostars and Planets IV, ed. V. Mannings, A. P. Boss \&
     S. S. Russell (Tucson: Univ. Arizona Press), 589
\bibitem[Verbunt \& Phinney(1995)]{ver95}
     Verbunt, F., \& Phinney, E. S. 1995, \aap, 296, 709
\bibitem[Ward(1997)]{war97}
     Ward, W. R. 1997, Icarus, 126, 261
\bibitem[Wisdom \& Holman(1991)]{wis91}
     Wisdom, J., \& Holman, M. 1991, \aj, 102, 1528
\bibitem[Yoder(1979)]{yod79}
     Yoder, C. F. 1979, \nat, 279, 747
\bibitem[Yoder \& Peale(1981)]{yod81}
     Yoder, C. F., \& Peale, S. J. 1981, Icarus, 47, 1
\bibitem[Zahn(1989)]{zah89}
     Zahn, J.-P. 1989, \aap, 220, 112
\end{thebibliography}
\end{document}